\documentclass[12pt,twoside,a4paper]{article}
\usepackage{amsmath,amssymb,latexsym,theorem,natbib,epsfig,color,subfigure,footmisc,bbm}
\usepackage{multirow,graphicx,array,rotating,url,multicol}  
\usepackage{epstopdf,afterpage,wasysym,hyperref}
\usepackage{verbatim}
\usepackage[normalem]{ulem}
\def\cov{\mathop{\hbox {\sf cov}}}

\numberwithin{equation}{section}

\newcommand{\degree}{\ensuremath{^\circ}}

\title{Fair Box ordinate transform for forecasts following a multivariate Gaussian law}

\author{S\'andor Baran$^{1,*}$ and Martin Leutbecher$^{2}$\\[0.5cm]
{\small $^1$Faculty of Informatics, University of Debrecen, Hungary}\\
{\small $^2$European Centre for Medium-Range Weather Forecasts, Reading, United Kingdom}
}

\date{}

\begin{document}
\maketitle

\footnotetext[1]{Corresponding author: \url{baran.sandor@inf.unideb.hu}}

\begin{abstract}
Monte Carlo techniques are the method of choice for making probabilistic predictions of an outcome in several disciplines. Usually, the aim is to generate calibrated predictions which are statistically indistinguishable from the outcome.

Developers and users of such Monte Carlo predictions are interested in evaluating the degree of calibration of the forecasts. Here, we consider predictions of $p$-dimensional outcomes sampling a multivariate Gaussian distribution and apply the Box ordinate transform (BOT) to assess calibration. However, this approach is known to fail to reliably indicate calibration when the sample size $n$ is moderate.

For some applications, the cost of obtaining Monte-Carlo estimates is significant, which can limit the sample size, for instance, in model development when the model is improved iteratively. Thus, it would be beneficial to be able to reliably assess calibration even if the sample size $n$ is moderate. To address this need, we introduce a fair, sample size- and dimension-dependent version of the Gaussian sample BOT.

In a simulation study, the fair Gaussian sample BOT is compared with alternative BOT versions for different miscalibrations and for different sample sizes. Results confirm that the fair Gaussian sample BOT is capable of correctly identifying miscalibration when the sample size is moderate in contrast to the alternative BOT versions.

Subsequently, the fair Gaussian sample BOT is applied to two to 12-dimensional  predictions of temperature and  vector wind using operational ensemble forecasts of the European Centre for Medium-Range Weather Forecasts (ECMWF). Firstly, perfectly reliable situations are considered where the outcome is replaced by a forecast that samples the same distribution as the members in the ensemble. Secondly, the BOT is computed using estimates of the actual temperature and vector wind from ECMWF analyses.

\end{abstract}

\section{Introduction}
\label{sec1}

In the last three decades, probabilistic forecasting using ensemble forecasts has become the standard approach in weather prediction \citep{lewis.2005,b18a}; all major weather centres operate their own ensemble prediction systems. Traditionally, these forecasts consist of multiple simulations with  a numerical weather prediction (NWP) model, each simulation using a different realisation for the initial uncertainties and model uncertainties \citep{leutbecher.palmer.2008}. The NWP model describes the evolution of the atmosphere and coupled earth system components based on the laws of physics with different initial conditions and/or model parametrizations. With recent rapid  advances in artificial intelligence, the methodology has been developed to  obtain data-driven weather forecast models suitable for ensemble forecasting by using machine learning \citep{lang2024aifs-crps,gencast25}.

In an ideal situation, ensemble forecasts would be perfectly reliable; that is, they would sample the same distribution as the future value of the weather quantity to be predicted. In this case, the relative frequency of an observed weather event would be statistically consistent with its forecast probability. This joint property of forecasts and observations is referred to as calibration. Unfortunately, ensemble forecasts still often suffer from the lack of calibration \citep[see e.g.][]{b18b}; nevertheless, this deficiency can be addressed by some form of statistical post-processing \citep{vbd21}, usually resulting in an estimate of the predictive distribution of the variable at hand.

There are several tools to assess the calibration of univariate probabilistic forecasts, which are probabilistic predictions of scalars, i.e.\  a single quantity for one location, one valid time and a single forecast lead time. In the case of ensemble predictions, one can think about, for instance, the verification rank histogram or Talagrand diagram \citep[][Section 9.7.1]{wilks.2019} or the reliability diagram \citep[][Section 9.4.4]{wilks.2019}. The latter can also be used for forecasts provided in the form of full predictive distributions. In  this case, it is appropriate to replace the rank histogram  by its continuous counterpart, the probability integral transform (PIT) histogram \citep[][Section 9.5.4]{wilks.2019}. Note, that besides assessing calibration, these simple graphical tools can also reveal the possible sources of unreliability. Furthermore, the level of calibration of different probabilistic forecasts can be compared with the help of proper scoring rules \citep{gneiting.raftery.2007}, which are loss functions assigning numerical values to forecast-observation pairs. However, in the case of ensemble predictions, as argued by \citet{ferro.ea.2008}, the ensemble size affects the score values, leading to the idea of fair scores \citep{fricker.ea.2013}. Later, \citet{ferro.2014} extended this notion and considered the fair versions of, for instance, the Brier score  \citep[][Section 9.4.2]{wilks.2019} or the continuous ranked probability score \citep[][Section 9.5.1]{wilks.2019}, while for Gaussian forecasts, \citet{siegert.ea.2019} determined the dependence of the logarithmic score \citep{good.1952} on the ensemble size and derived its fair version.

Nowadays, more and more applications require multivariate predictions, which can comprise joint forecasts of several weather quantities such as visibility and cloud ceiling, jointly determining the low-visibility procedure states at airports \citep{dkmz19}, predictions of the same scalar variable for consecutive time points, having importance in hydrological forecasting \citep{hlk.2015}, or for various locations, having been utilized, for instance, in winter road maintenance \citep{brgs10}, or any combination of the cases above \citep{schefzik.2016a}.

The most popular graphical tools for multivariate calibration assessment of ensemble forecasts are the multivariate rank histograms  \citep{gneiting.ea.2008}, which extend the idea of the Talagram diagram and are based on the concept of pre-ranks. For detailed comparisons of various pre-rank functions, we refer to \citet{tsh16} and \citet{azg24}. Furthermore, for continuous probabilistic forecasts represented by a multivariate predictive distribution, \citet{gneiting.ea.2008} propose the application of the Box ordinate transform (BOT). In a detailed simulation study focusing on multivariate Gaussian predictions, \citet{wilks.2017} compares the performance of the rank histograms studied by \citet{tsh16} and the Gaussian sample BOT in identifying various forms of calibration misspecification. For the latter, the mean vector and the covariance matrix of the Gaussian predictive distribution are estimated using both the simulated forecast and the corresponding observation. However, this approach is based on the approximate distribution of the Gaussian BOT and performs well only if the ensemble size is substantially larger than the dimension of the forecast; for instance, in the three-dimensional case, the suggested ensemble size is about $100$.

In the present work, following some of the ideas of \citet{lb25} for determining the ensemble size dependence of the logarithmic score under the assumption of multivariate normality, we determine the fair version of the Gaussian sample BOT. At first, the behaviour of the proposed fair sample BOT is studied for simulated forecast-observation pairs in the case of reliable forecasts and in  cases of forecasts with mispecified covariance or mispecified mean. The fair sample BOT is compared to the characteristics of the theoretical BOT and two different Gaussian sample BOT versions based on distinct estimates of the mean vector and the covariance matrix of the predictive law. Then, we consider multivariate 2-, 4-, 9-, and 12-dimensional examples based on operational ECMWF subseasonal forecasts of temperature and vector wind (i) in a perfectly calibrated setup when one of the ensemble members plays the role of the analysis and (ii) in an actual verification setting where the forecasts are verified against analyses.    

The paper is organized as follows. In Section \ref{sec2}, we derive the fair version of the Gaussian sample BOT. The simulation setup and the results of the various simulations are reported in Section \ref{sec3}. Section \ref{sec4} contains the verification results based on operational ensemble forecasts, followed by discussion and conclusions in Section \ref{sec5}. Finally, further simulation results and operational forecast-based verification results can be found in Sections \ref{secA} and \ref{secB} of the Appendix, respectively.

\section{The fair Gaussian sample Box ordinate transform}
\label{sec2}
Assume, that the predictive distribution for a $p$-dimensional quantity materialized in observation \ $\boldsymbol x_0$ \ is a $p$-dimensional Gaussian distribution  \ ${\mathcal N}_p\big(\boldsymbol\mu, \boldsymbol\Sigma\big)$ \ with mean vector \ $\boldsymbol \mu$ \ and regular covariance matrix \ $\boldsymbol\Sigma$. \ In this particular case the BOT has the form
\begin{equation}
  \label{eq:botTh}
u\big(\boldsymbol\mu,\boldsymbol\Sigma;\boldsymbol x_0\big)=1 - {\mathcal X}^2_p\Big[\big(\boldsymbol x_0 - \boldsymbol\mu\big)^{\top} {\boldsymbol\Sigma}^{-1}\big(\boldsymbol x_0 - \boldsymbol\mu\big)\Big],
\end{equation}
where \ ${\mathcal X}^2_p[\,\cdot\,]$ \ denotes the cumulative distribution function (CDF) of a chi-square distribution with \ $p$ \ degrees of freedom \citep{gneiting.ea.2008}. If the predictive distribution matches the distribution of the observation, that is \ $\boldsymbol x_0 \sim {\mathcal N}_p\big(\boldsymbol\mu, \boldsymbol\Sigma\big)$, \ then the distribution of \ $u$ \ is standard uniform. In fact, in this case \ $1-u$ \ is standard uniform as well; however, to be consistent with the notation used in earlier studies \citep{gneiting.ea.2008, wilks.2017}, we prefer the form \eqref{eq:botTh}. 

Now, let the Gaussian predictive distribution be represented by an $n$-member forecast ensemble \ $\boldsymbol x_1,\boldsymbol x_2, \ldots ,\boldsymbol x_n$ \ comprising an independent sample drawn from \ ${\mathcal N}_p\big(\boldsymbol\mu, \boldsymbol\Sigma\big)$. \ In this case, a natural approach is to replace the population mean vector  \ $\boldsymbol\mu$ \ and population covariance matrix \ $\boldsymbol{\Sigma}$ \ by their unbiased estimators, that is by the sample mean \ $\boldsymbol m$ \ and sample covariance matrix \ $\boldsymbol{\mathsf S}$, \ respectively, defined as
\begin{equation}
  \label{eq:ensPars}
  {\boldsymbol m}
  :=\frac 1n\sum_{i=1}^n \boldsymbol x_i \qquad  \text{and} \qquad \boldsymbol{\mathsf S}:=\frac 1{n-1}\sum_{i=1}^n\big(\boldsymbol x_i-{\boldsymbol m}\big)\big(\boldsymbol x_i-{\boldsymbol m}\big)^{\top}.
\end{equation}
Note that for regular \ $\boldsymbol\Sigma$ \ and \ $n>p$ \ the corresponding sample covariance matrix  \ $\boldsymbol{\mathsf S}$ \ is also regular with probability 1.

However, when the ensemble members and the observation are drawn from the same Gaussian law, the distribution of the naive sample BOT \ $u\big(\boldsymbol m,\boldsymbol{\mathsf S};\boldsymbol x_0\big)$ \ is only asymptotically uniform and for small ensemble sizes \ $n$ \ the deviation can be substantial. The same applies to the adjusted version \ $u\big(\widetilde{\boldsymbol m},\widetilde{\boldsymbol{\mathsf S}};\boldsymbol x_0\big)$ \ suggested by \citet{wilks.2017}, where
\begin{equation}
  \label{eq:adjPars}
  \widetilde{{\boldsymbol m}}
  :=\frac 1{n+1}\sum_{i=0}^n \boldsymbol x_i \qquad  \text{and} \qquad \widetilde{\boldsymbol{\mathsf S}}:=\frac 1n\sum_{i=0}^n\big(\boldsymbol x_i-\widetilde{\boldsymbol m}\big)\big(\boldsymbol x_i-\widetilde{\boldsymbol m}\big)^{\top},
\end{equation}
that is the observation \ $\boldsymbol x_0$ \ is also used to estimate the population mean vector and covariance matrix. 

We aim to derive a fair sample version \ $u_n^F\big(\boldsymbol m,\boldsymbol{\mathsf S};\boldsymbol x_0\big)$ \ of the BOT which for calibrated forecasts follows a standard uniform law for any ensemble size \ $n$. \

Consider the squared Mahalanobis distance of  the observation vector \ $\boldsymbol x_0$ \ and the ensemble mean \ $\boldsymbol m$
\begin{equation*}
  {\mathcal D}^2:=\big(\boldsymbol x_0 - {\boldsymbol m}\big)^{\top}\boldsymbol{\mathsf S}^{-1}\big(\boldsymbol x_0 - {\boldsymbol m}\big).
\end{equation*}
According e.g. to \citet[][Theorem 5.7]{hardle.simar.2019}, \ ${\boldsymbol m}$ \ and \ $\boldsymbol{\mathsf S}$ \ are independent, \ ${\boldsymbol m} \sim  {\mathcal N}_p\big(\boldsymbol\mu, n^{-1}\boldsymbol\Sigma\big)$, \ whereas \ $(n-1)\boldsymbol{\mathsf S}$ \ follows a $p$-dimensional Wishart distribution \ $\mathcal W_p(n-1,\boldsymbol\Sigma)$ \ with \ $n-1$ \ degrees of freedom and scale matrix \ $\boldsymbol\Sigma$. \ Obviously, the observation vector \ $\boldsymbol x_0 $ \ is also independent both from  \ ${\boldsymbol m}$ \ and \ $\boldsymbol{\mathsf S}$. \ A short straightforward calculation verifies that if \ ${\boldsymbol x}_0 \sim  {\mathcal N}_p\big(\boldsymbol\mu, \boldsymbol\Sigma\big)$, \ then 
\begin{equation*}
 {\mathsf E}\big(\boldsymbol x_0 - {\boldsymbol m}\big) = \boldsymbol 0 \qquad \text{and} \qquad {\mathsf E}\big(\boldsymbol x_0 - {\boldsymbol m}\big)\big(\boldsymbol x_0 - {\boldsymbol m}\big)^{\top} = \frac{n+1}n\boldsymbol\Sigma.
  \end{equation*}
  Hence,
  \begin{equation*}
    \boldsymbol Y:= \sqrt{\frac n{n+1}}\big(\boldsymbol x_0 - {\boldsymbol m}\big)  \sim  {\mathcal N}_p\big(\boldsymbol 0,\boldsymbol\Sigma\big)
  \end{equation*}
and independent of \ $\boldsymbol{\mathsf S}$. \ In this way the rescaled squared Mahalanobis distance of the observation vector and the sample mean
  \begin{equation*}
    T^2:=\frac n{n+1}{\mathcal D}^2=\frac n{n+1}\big(\boldsymbol x_0 - {\boldsymbol m}\big)^{\top}\boldsymbol{\mathsf S}^{-1}\big(\boldsymbol x_0 - {\boldsymbol m}\big)
  \end{equation*}
  follows a $p$-variate Hotelling's $T^2$-distribution \ $T^2_{p,n-1}$ \ with \ $(n-1)$ \ degrees of freedom \citep[][Corollary 5.3]{hardle.simar.2019}. Furthermore, as the $T^2$-distribution \ $T^2_{p,n-1}$ \ is proportional to the $F$-distribution \ $F_{p,n-p}$ \ with \ $p$ \ and \ $n-p$ \ degrees of freedom \citep[][Theorem 5.9]{hardle.simar.2019}, one has
  \begin{equation}
    \label{eq:distrT2}
    \frac{n-p}{p(n-1)}T^2 = \frac{n(n-p)}{p(n^2-1)}\big(\boldsymbol x_0 - {\boldsymbol m}\big)^{\top}\boldsymbol{\mathsf S}^{-1}\big(\boldsymbol x_0 - {\boldsymbol m}\big) \sim F_{p,n-p}.
  \end{equation}

  By \eqref{eq:distrT2}, the formula for the fair sample BOT is
  \begin{equation}
    \label{eq:fairBOT}
    u_n^F\big(\boldsymbol m,\boldsymbol{\mathsf S};\boldsymbol x_0\big) =
    1 - {\mathcal F}_{p,n-p}\left[\frac{n(n-p)}{p(n^2-1)}\big(\boldsymbol x_0 - {\boldsymbol m}\big)^{\top}\boldsymbol{\mathsf S}^{-1}\big(\boldsymbol x_0 - {\boldsymbol m}\big) \right],
  \end{equation}
where \ ${\mathcal F}_{p,n-p}[\,\cdot\,]$ \ denotes the CDF of an $F$-distribution with \ $p$ \ and \ $n-p$ \ degrees of freedom.

\section{Simulation study}
\label{sec3}

First, following the experimental setup of \citet{wilks.2017}, we explore the behaviour of the fair sample BOT \eqref{eq:fairBOT} in a simulation study, where various dimensions \ $p$, \ ensemble sizes \ $n$, \ and types of calibration misspecification are considered.

\subsection{Model configuration}
\label{subs3.1}

Let the observation vector be centred $p$-dimensional Gaussian with covariance matrix \ $\boldsymbol\Sigma _0$, \ that is
\begin{equation*}
  \boldsymbol x_0 = \big(x_{0,1},x_{0,2}, \ldots ,x_{0,p}\big)^{\top} \sim {\mathcal N}_p\big(\boldsymbol 0,\boldsymbol\Sigma_0\big).
  \end{equation*}
The entries of \ $\boldsymbol\Sigma_0$, \ which provide the ground truth in terms of the dependence structure of the coordinates are
\begin{equation}
  \label{eq:trueCov}
  \cov\big(x_{0,k},x_{0,\ell}\big)=\sigma_0^2\varrho_0 ^{|k-\ell|}, \qquad k,\ell = 1,2, \ldots ,p.
\end{equation}  
Note that \ $\boldsymbol\Sigma _0$ \ is the \ $p\times p$ \ autocovariance matrix of a stationary first-order autoregressive (AR(1)) process
\begin{equation}
  \label{eq:ar1}
  x_t = \varrho_0 x_{t-1} + \varepsilon _t, \qquad \text{where} \qquad \varepsilon_t \sim {\mathcal N}_1\big(0,\sigma_0^2(1-\varrho_0^2)\big),
\end{equation}
which can serve as a simple model for daily minimum or maximum temperature, see e.g. \citet{richardson.1981} or \citet[][Section 10.3.1]{wilks.2019}. Hence, one can consider \ $\boldsymbol x_0$ \ as the vector of temperature observations for \ $p$ \ consecutive calendar days.

The simulated forecast ensemble is an independent sequence
\begin{equation*}
  \boldsymbol x_i =  \big(x_{i,1},x_{i,2}, \ldots ,x_{i,p}\big)^{\top} \sim {\mathcal N}_p\big(\boldsymbol \mu,\boldsymbol\Sigma_f\big), \qquad i=1,2,\ldots ,n,
\end{equation*}
which is calibrated if \ $\boldsymbol\mu = \boldsymbol 0$ \ and \ $\boldsymbol\Sigma_f=\boldsymbol\Sigma_0$. \ 

First, we assume that \ $\boldsymbol\Sigma_f$ \ has the same structure as \ $\boldsymbol\Sigma_0$ \ with entries
\begin{equation}
  \label{eq:ensCov}
  \cov\big(x_{i,k},x_{i,\ell}\big)=\sigma_f^2\varrho_f ^{|k-\ell|}, \qquad k,\ell = 1,2, \ldots ,p, \quad i=1,2,\ldots ,n.
\end{equation}

Additionally, we consider a more complex covariance structure 
\begin{equation}
  \label{eq:ensCov2}
  \cov\big(x_{i,k},x_{i,\ell}\big)=\sqrt{\sigma_0^2 + (-1)^k\sigma_\Delta^2}\sqrt{\sigma_0^2 + (-1)^\ell\sigma_\Delta^2}\left(\varrho_0+(-1)^{|k-\ell|}\varrho_\Delta\right)^{|k-\ell|}, 
\end{equation}
$k,\ell = 1,2, \ldots ,p$, \ $i=1,2,\ldots ,n$, \ where the true variance \ $\sigma_0^2$ \ and correlation \ $\varrho$ \ are loaded with additive errors \ $\sigma_\Delta^2$ \ and \ $\varrho_\Delta$, \ respectively, having alternating signs.

In our simulation study, we investigate the existence of bias in the forecasts \ ($\boldsymbol\mu \ne \boldsymbol 0$) \ and the incorrect specification of the covariance structure \ ($\boldsymbol\Sigma_f\ne\boldsymbol\Sigma_0$) \  separately. In the latter case for covariance structure \eqref{eq:ensCov}, first we consider the situations \ $\sigma_f^2\ne \sigma_0^2, \ \varrho_f=\varrho_0$ \ (misspecified marginal variances, correct correlations) and  \ $\sigma_f^2= \sigma_0^2, \ \varrho_f\ne\varrho_0$ \ (correct marginal variances, misspecified correlations) referred to by  \citet{wilks.2017} as Type 1 and Type 2 covariance miscalibrations, respectively. Then, cases when both the marginal variances and the correlations are misspecified are studied. Finally, in the case of covariance structure \eqref{eq:ensCov2}, we again treat the errors in marginal variances \ ($\sigma_\Delta^2>0, \ \varrho_\Delta =0$) \ and the correlation errors \ ($\sigma_\Delta^2=0, \ \varrho_\Delta >0$) \ separately.

\subsection{Simulation results}
\label{subs3.2}

\begin{figure}[!h]
   \centering
   \epsfig{file=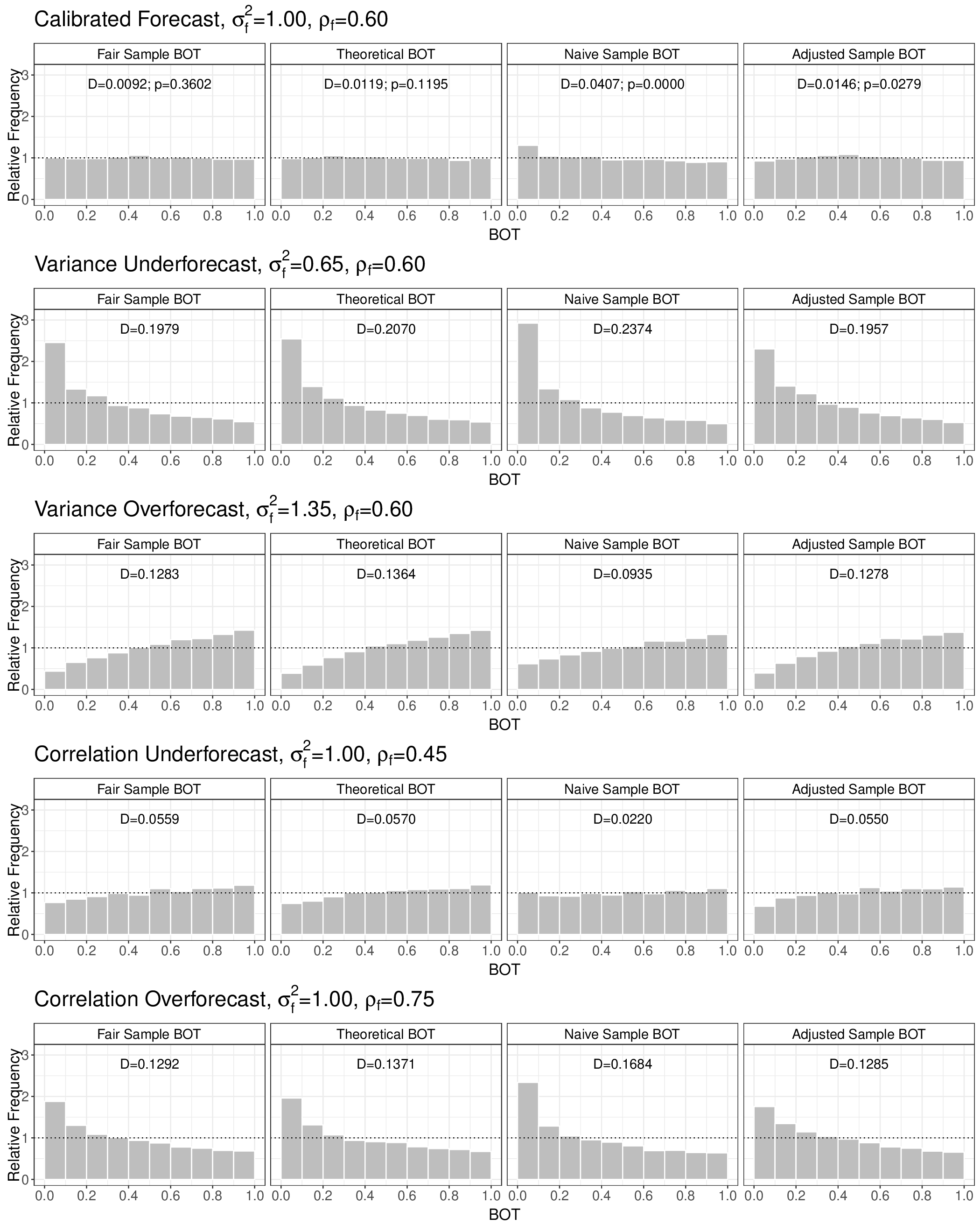, width=.98\textwidth} 
   \caption{Histograms of various BOT versions for calibrated 3-dimensional 50-member ensemble forecasts and misspecified predictions together with the test statistic \ $D$ \ of the Kolmogorov-Smirnov test for uniformity. For calibrated forecasts, the corresponding $p$-values are also provided.}
   \label{fig:BOTd3n50}
 \end{figure}

In each of the following cases we generate 10000 ensemble forecasts and examine the shape of histograms of the fair sample BOT values \eqref{eq:fairBOT}. As references we also display the corresponding histograms of the theoretical BOT \eqref{eq:botTh}, the naive sample BOT based on the ensemble mean and ensemble covariance matrix \eqref{eq:ensPars}, and its adjusted version relying on parameter estimates \eqref{eq:adjPars} incorporating the observation as well. We also report the values of the test statistic \ $D$ \ of the Kolmogorov-Smirnov test for uniformity of the BOT values, and for calibrated forecasts, we provide the corresponding $p$-values as well. Following \citet{wilks.2017}, we set \ $\sigma_0^2=1$ \ and \ $\varrho_0=0.6$. In Section \ref{subs3.2.1}, results for the calibrated predictions will be discussed together with the miscalibrated ones.

\subsubsection{Misspecified covariance structure}
\label{subs3.2.1}
We assume that the ensemble forecasts are unbiased \ ($\boldsymbol \mu=\boldsymbol 0$) \ and first consider four different cases of misspecification of the covariance structure \eqref{eq:ensCov}. Subject to correct correlations \ ($\varrho_f=\varrho_0=0.6$), \ we set \ $\sigma_f^2=0.65$ \ (variance underforecast) and \ $\sigma_f^2=1.35$ \ (variance overforecast), whereas in the case of correct variances \ ($\sigma_f^2=\sigma_0^2=1$) \ we consider \ $\varrho_f=0.45$ \ (correlation underforecast) and \ $\varrho_f=0.75$ \ (correlation overforecast), these values are the same as in \citet{wilks.2017}.

\begin{figure}[!h]
 \centering
   \epsfig{file=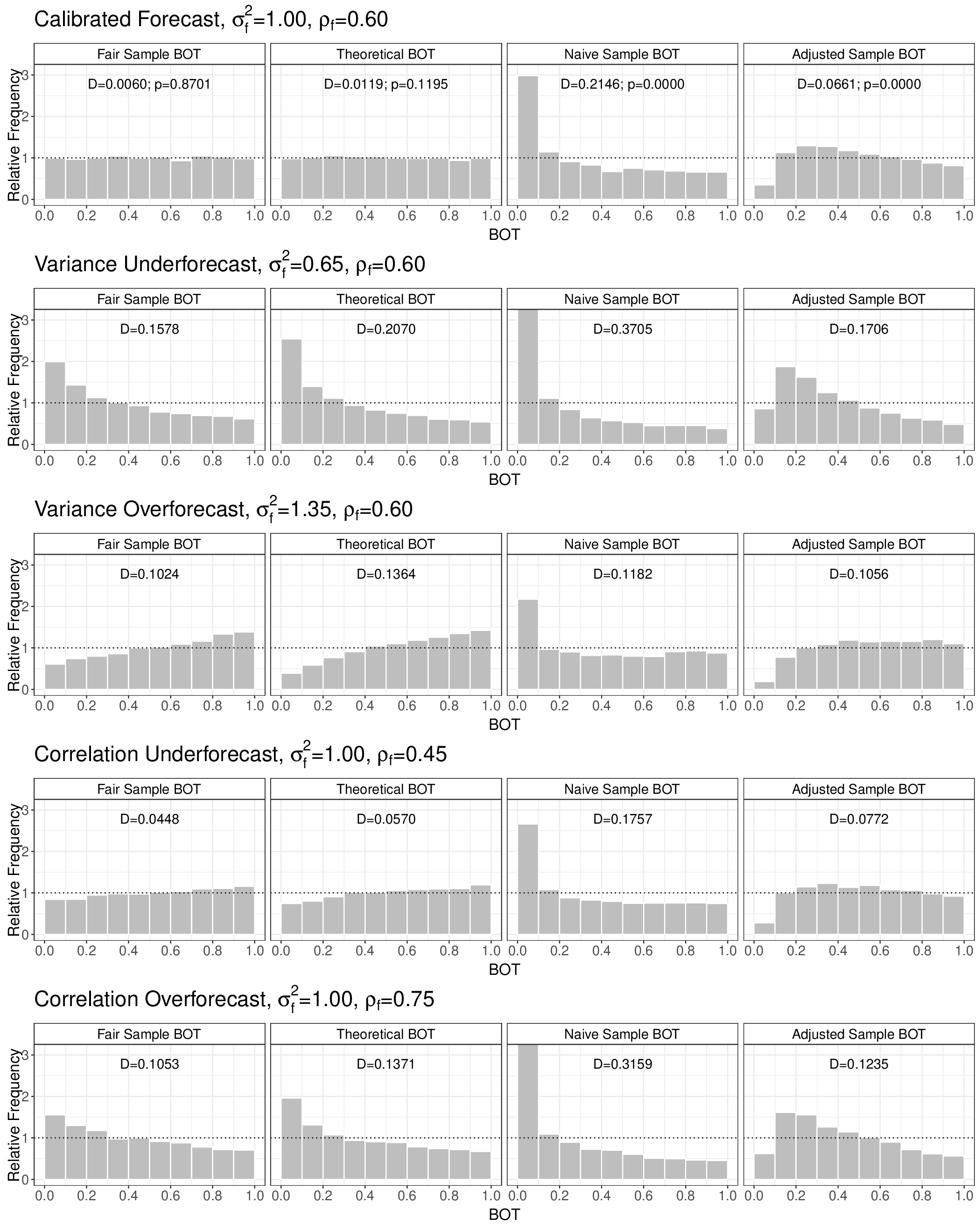, width=.98\textwidth} 
   \caption{Histograms of various BOT versions for calibrated 3-dimensional 10-member ensemble forecasts and misspecified predictions together with the test statistic \ $D$ \ of the Kolmogorov-Smirnov test for uniformity. For calibrated forecasts, the corresponding $p$-values are also provided.}
   \label{fig:BOTd3n10}
 \end{figure}

Figure \ref{fig:BOTd3n50} shows the various BOT histograms for 3-dimensional 50-member ensemble forecasts. In the calibrated case, the fair sample BOT and the theoretical BOT result in completely flat histograms with $p$-values of $0.3602$ and $0.1195$, respectively, confirming uniformity. In contrast, the naive sample BOT histogram is slightly biased in the direction of lower bins ($p<5\times 10^{-5}$), while the histogram of the adjusted sample BOT is a bit hump-shaped, yielding a low $p$-value of $0.0279$. The responses to Type 1 miscalibration are rather consistent for fair sample, theoretical, and adjusted sample BOTs, with \ $D$ \ values around $0.2$ for the underforecast variance and around $0.13$ for the overforecast variance case. The corresponding naive sample BOT histograms show similar shapes as the other three BOT versions: heavily right skewed for \ $\sigma^2_f < \sigma^2_0$ \ and displaying a linear increase towards the higher bins  for \ $\sigma^2_f > \sigma^2_0$. However, the deviation from the uniformity of the naive sample BOT differs from the deviations seen for the other BOT versions. It is greater when the variance is over-forecasted ($D=0.2374$) and smaller when it is under-forecasted ($D=0.0935$). The situation is reversed
in the case of Type 2 miscalibration. Overforecast correlation is characterized by strongly right-skewed histograms, while underforecast correlation results in slight overpopulation of the higher bins. Again, the naive sample BOT shows slightly different behaviour as the corresponding histograms are almost flat for under-forecasted correlation and more skewed than the others when the correlation is over-forecasted. Note that the shapes of the adjusted sample BOT histograms are consistent with those of the corresponding histograms in Figure 1 of \citet{wilks.2017}.

\begin{figure}[!h]
   \centering
   \epsfig{file=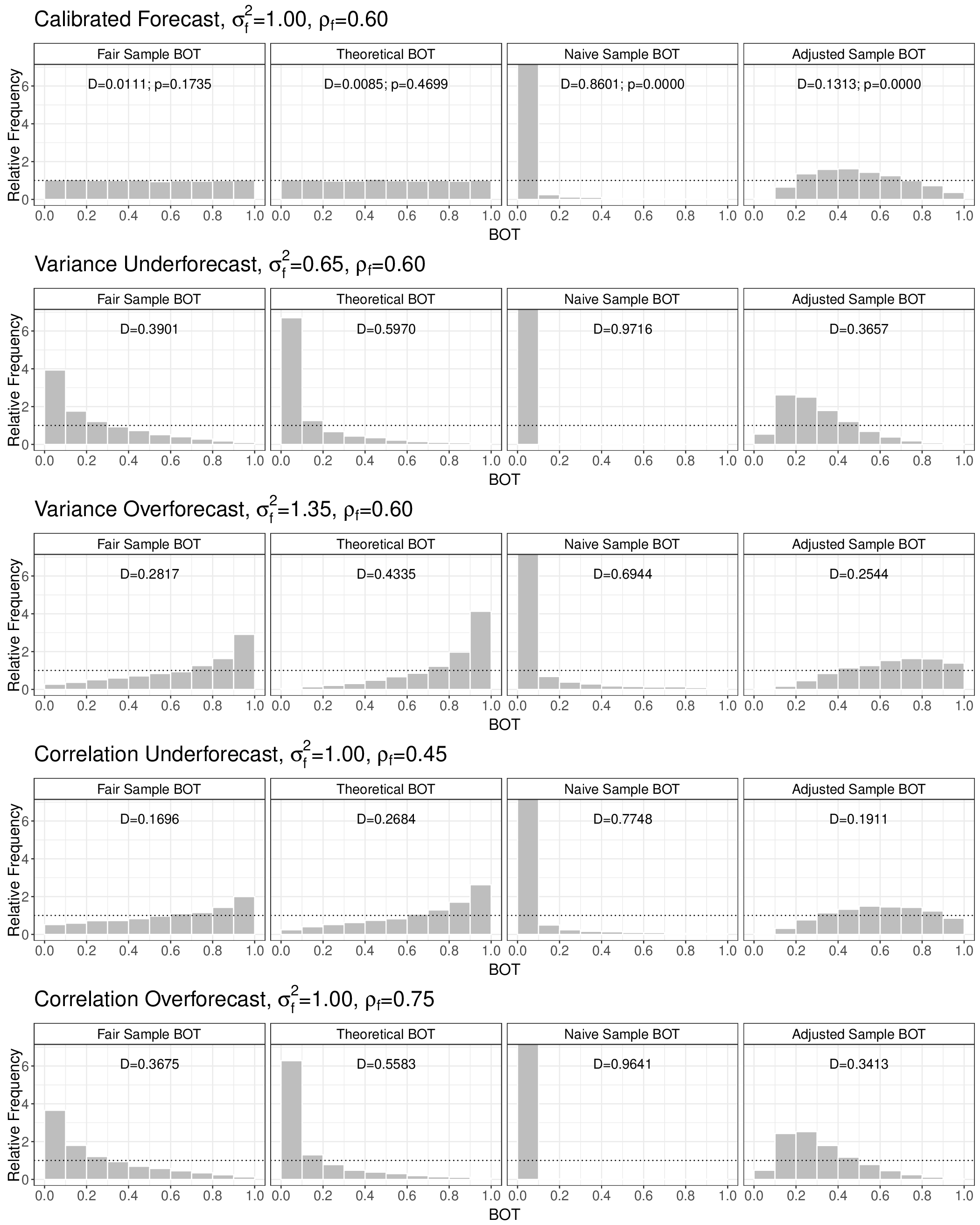, width=.98\textwidth} 
   \caption{Histograms of various BOT versions for calibrated 30-dimensional 50-member ensemble forecasts and misspecified predictions together with the test statistic \ $D$ \ of the Kolmogorov-Smirnov test for uniformity. For calibrated forecasts, the corresponding $p$-values are also provided.}
   \label{fig:BOTd30n50}
 \end{figure}

The reduction of the ensemble size, as depicted in Figure \ref{fig:BOTd3n10}, drastically changes the situation. While the shapes of the fair sample BOT histograms of 10-member forecasts are the same as their 50-member counterparts -- in the calibrated case, uniformity is accepted on a very high level of significance ($p=0.8701$) -- all naive sample BOT histograms are heavily right-skewed, whereas, in all adjusted sample BOT histograms, the lower bins are underpopulated, resulting in hump-shapes. However, one should also note that the differences between the matching fair sample BOT and theoretical BOT histograms (which do not depend on the ensemble size) are greater than what can be observed in Figure \ref{fig:BOTd3n50}; the BOT values of the former version are consistently closer to the uniform distribution in all studied cases than those of the latter.

 \begin{figure}[t]
   \centering
   \epsfig{file=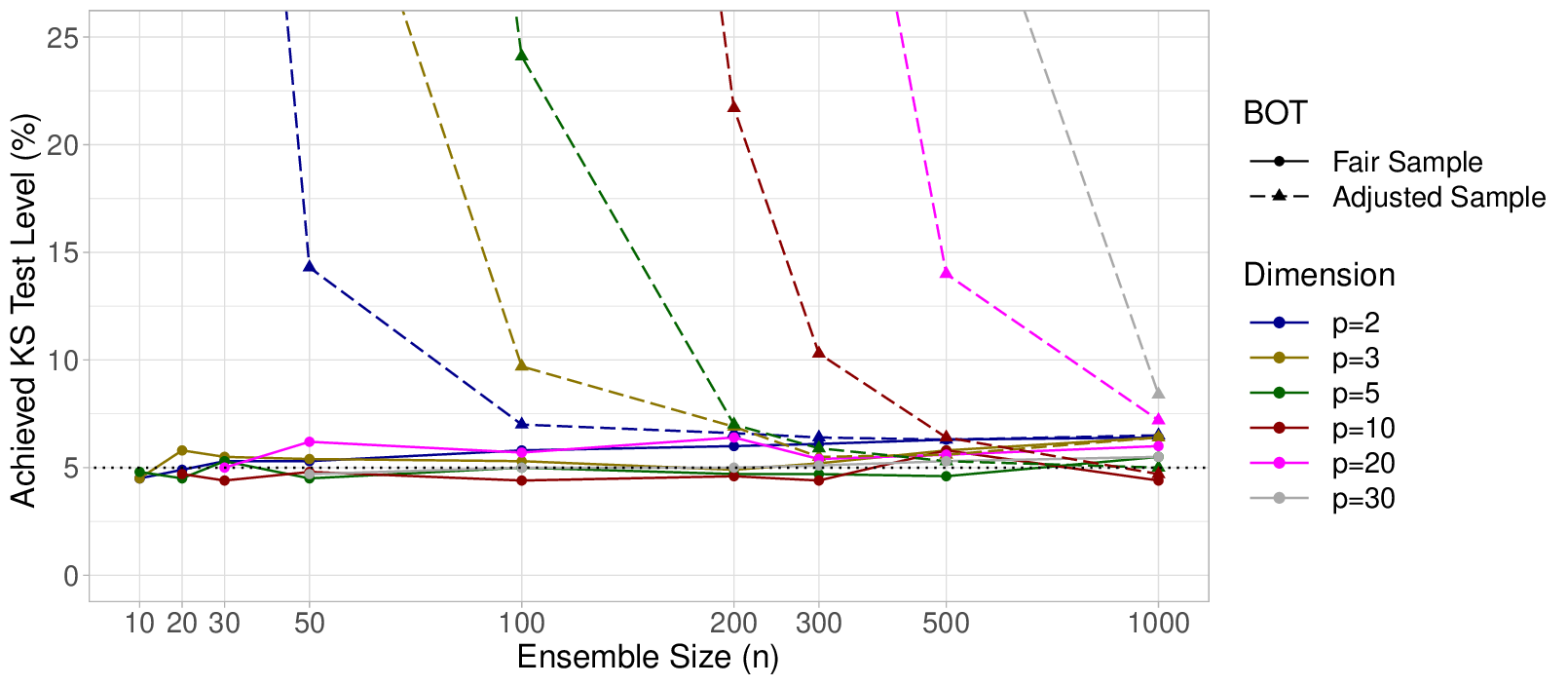, width=.75\textwidth} 
   \caption{Proportions of Kolmogorov-Smirnov tests (1000 simulations) rejecting uniformity of the BOT values at a 5\,\% level for properly calibrated forecasts for various dimensions \ $p$ \ as functions of the ensemble size \ $n$.}
   \label{fig:KSlevels}
 \end{figure}

 The inferior behaviour of the naive and adjusted sample BOT values is even more pronounced in Figure \ref{fig:BOTd30n50}, where 30-dimensional, 50-member ensemble forecasts are considered. The naive sample BOT puts almost all weight on the lowest bin in all cases, and in this way, it is completely useless. The skewness directions of the different adjusted sample BOT histograms resemble the correct directions indicated by the corresponding theoretical ones; however, in this case, both the highest and the lowest bins are strongly underpopulated. In contrast, the fair sample BOT performs well even in this situation. The shapes of the histograms in the misspecified cases mimic those of the corresponding theoretical BOT histograms, while for calibrated forecasts the uniformity is accepted on a convincing significance level ($p=0.1735$).

\begin{figure}[t]
   \centering
   \epsfig{file=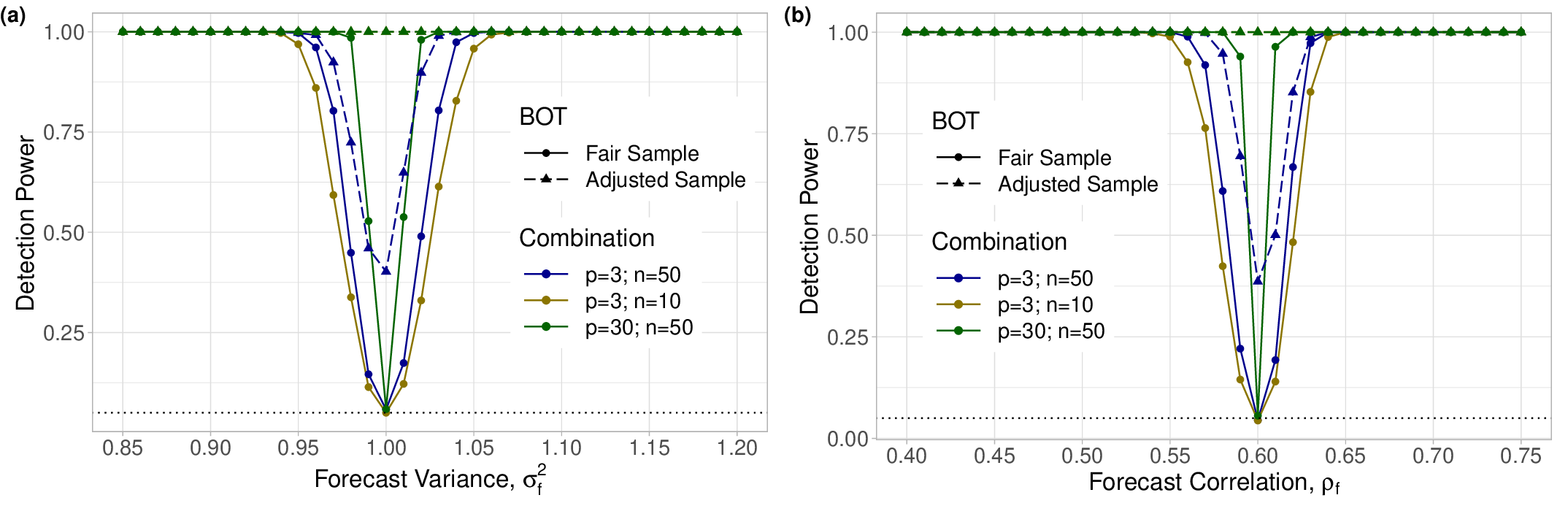, width=\textwidth} 
   \caption{Power curves of Kolmogorov-Smirnov tests (1000 simulations) for detecting (a) Type 1 and (b) Type 2 miscalibration at a 5\,\% level of significance.}
   \label{fig:KSpower}
 \end{figure} 
 
 Figures \ref{fig:BOTd3n50} -- \ref{fig:BOTd30n50} confirm the findings of \citet{wilks.2017} that the adjusted sample BOT performs well if \ $n\gg p$, \ whereas the fair sample BOT does not require this constraint. The same conclusion can be drawn from Figure \ref{fig:KSlevels} showing the proportion of Kolmogorov-Smirnov tests in 1000 simulations (each consisting of 10000 ensembles) rejecting uniformity of the fair and adjusted sample BOT values at a 5\,\% level for properly calibrated forecasts. For the fair sample BOT, all values oscillate around the desired 5\,\% regardless of the dimension \ $p$ \ and ensemble size \ $n$. \ In contrast, the test levels for the adjusted sample BOT approximate 5\,\% at \ $n\approx 300$ \ for 2-, 3-, and 5-dimensional and \ $n\approx 1000$ \ for 10-dimensional forecasts, whereas with dimensions 20 and 30 even this ensemble size appears still too small.  

 \begin{figure}[!th]
 \centering
   \epsfig{file=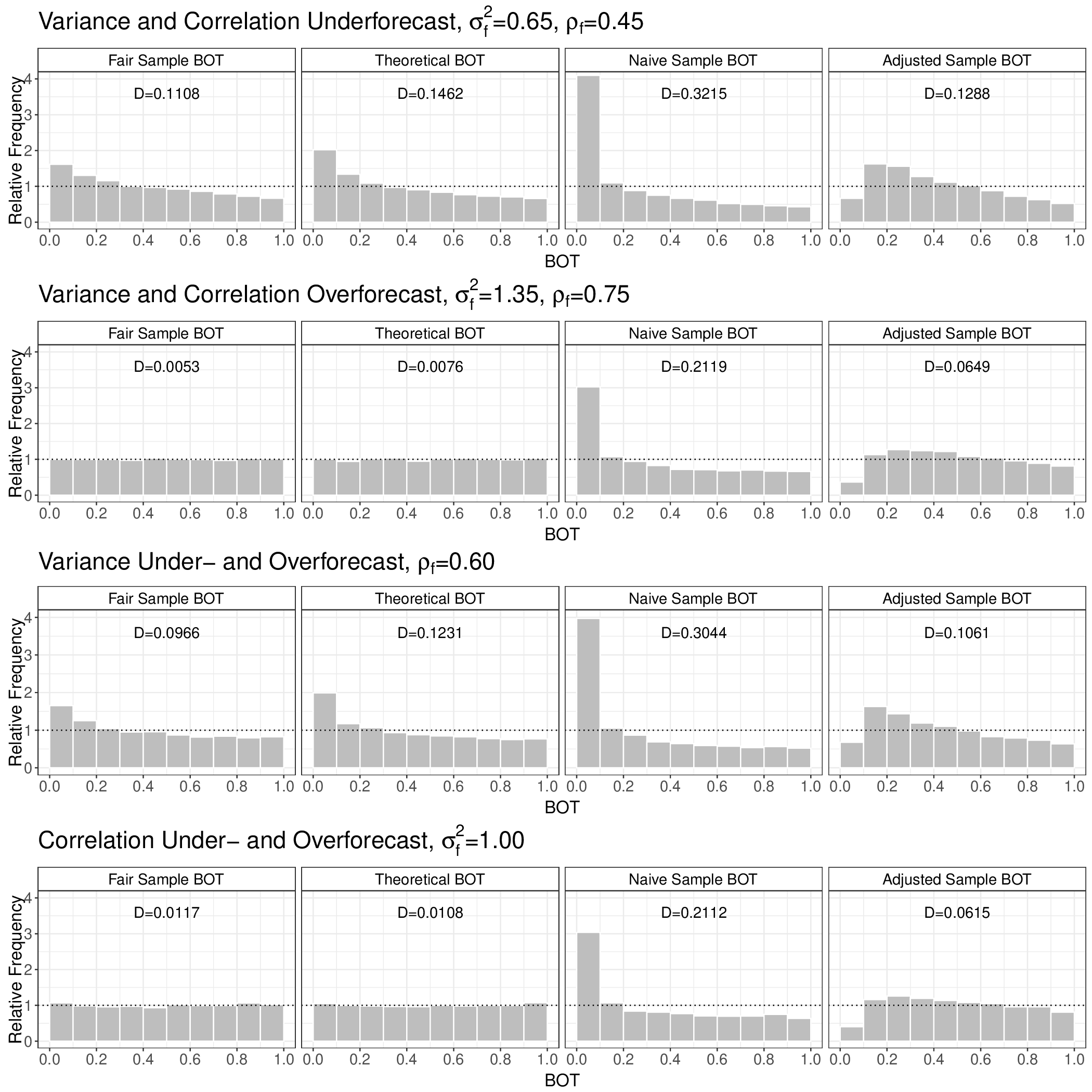, width=.98\textwidth} 
   \caption{Histograms of various BOT versions for misspecified 3-dimensional 10-member ensemble forecasts together with the test statistic \ $D$ \ of the Kolmogorov-Smirnov test for uniformity.}
   \label{fig:BOTmixd3n10}
 \end{figure}

 Finally, in Figure \ref{fig:KSpower} the power curves of Kolmogorov-Smirnov tests providing the estimated probabilities (from 1000 simulations) of rejecting the null hypothesis of uniformity of the BOT values at a 5\,\% level for various alternatives are depicted. In the case of Type 1 miscalibration the forecast variance \ $\sigma^2_f$ \ ranges from 0.85 to 1.2 (Figure \ref{fig:KSpower}a), whereas for Type 2 miscalibration, forecast correlation \ $\varrho_f$ \ varies in the interval $[0.4,0.75]$ (Figure \ref{fig:KSpower}b). The fair sample BOT results in the usual $\cup$-shaped power curves in all investigated cases, the ranking of the various combinations is the same for \ $\sigma^2_f$ \ and \ $\varrho_f$. \
 In contrast, the adjusted sample BOT demonstrates some power only if  \ $n\gg p$ (case \ $n=50, \ p=3$); however, for calibrated forecasts it fails to reach the desired 5\,\%. For the other two combinations, the Kolmogorov-Smirnov test rejects the uniformity of the adjusted sample BOT values in all cases for all investigated values of the forecast variance and forecast correlation (even for calibrated forecasts). In this way, the power curves of the adjusted sample BOT are in sharp contrast to the corresponding curves of Figures 2a,b of \citet{wilks.2017}. Note, that in \citet{wilks.2017} \ $\sigma^2_f$ \ and \ $\varrho_f$ \ vary in intervals \ $[0.6,1.6]$ \ and \ $[0,0.9]$, \ respectively, and the sensitivity of the chi-square test for uniformity for the adjusted sample BOT is weak for small correlations.  Considering the same intervals in our case, outside the segments displayed in Figures \ref{fig:KSpower}a,b all power values of the Kolmogorov-Smirnov test are identically 1.

Moving to more complex forms of misscalibration, for the covariance structure \eqref{eq:ensCov}, we also study the cases when the variance and the correlation are simultaneously either underforecast \ ($\sigma_f^2=0.65, \ \varrho_f=0.45$) \ or overforecast \ ($\sigma_f^2=1.35, \ \varrho_f=0.75$). \ Additionally,  we also perform simulations with covariances defined by \eqref{eq:ensCov2}; however, in this case, misspecified marginal variances and misspecified covariance are again treated separately. First, we assume the correct correlation \ ($\varrho_\Delta=0, \ \varrho_f=\varrho_0=0.6$) \ and set \ $\sigma^2_\Delta=0.35$; \ that is, the marginal variances alternate between $0.65$ and $1.35$ (variance under- and overforecast). Then, the marginal variances are specified correctly \ ($\sigma^2_\Delta=0, \ \sigma^2_f=\sigma^2_0=1$), \ and the correlation alternates between $0.45$ and $0.75$, so \ $\varrho_\Delta=0.15$. We refer to this case as correlation under- and overforecast.

For 3-dimensional 50-member ensemble forecasts, similar to Type 1 and Type 2 miscalibrations displayed in Figure \ref{fig:BOTd3n50}, the histograms of the fair sample, theoretical, and adjusted sample BOT values are very similar in each of the four investigated types of misspecified covariance structure, and the corresponding values \ $D$ \ of the Kolmogorov-Smirnov test statistic are also close. Only the naive sample BOT exhibits a stronger bias towards the lower bins, resulting in higher \ $D$ \ values than its competitors (see Figure \ref{fig:BOTmixd3n50} in Section \ref{subsA.1} of the Appendix).

As depicted in Figure \ref{fig:BOTmixd3n10}, the reduction of the ensemble size to dimension ratio \ $n/p$ \ results in even more biased naive sample BOT values and also has an inferior effect on the performance of the adjusted sample BOT. The histograms of the latter are now hump-shaped with strongly underpopulated lower bins in each studied case, whereas the shapes of the fair sample BOT histograms mimic the forms of their theoretical counterparts. Note that in configurations when both marginal variances and correlation are overforecast and when the marginal variances are correct, but the correlations are under- and overforecast, despite the serious miscalibration, the fair sample and theoretical BOT histograms are almost flat. For \ $n=50$ \ and \ $p=3$, \ the same applies to the adjusted sample BOT; see Figure \ref{fig:BOTmixd3n50} in Section \ref{subsA.1} of the Appendix.  This behaviour is the result of some kind of compensation since, for instance, according to Figure \ref{fig:BOTd3n10}, variance overforecast results in right-skewed fair sample and theoretical BOT histograms, whereas the corresponding histograms in the correlation overforecast case are left-skewed. To avoid being misled by the flat fair sample BOT histograms suggesting proper calibration, on the one hand, one should combine the study of the BOT with tests for the reliability of the marginal distributions. This extension will handle the case when both the marginal variances and the correlation are overforecast. On the other hand, the correlation under and overforecast can be captured by investigating the fair sample BOT histograms of the prediction corresponding to various lower-dimensional subspaces.

Further histograms corresponding to 30-dimensional 50-member forecasts can be found in Section \ref{subsA.1} of the Appendix (Figure \ref{fig:BOTmixd30n50}).
 
 \subsubsection{Bias}
 \label{subs3.2.2}

 \begin{figure}[th!]
   \centering
   \epsfig{file=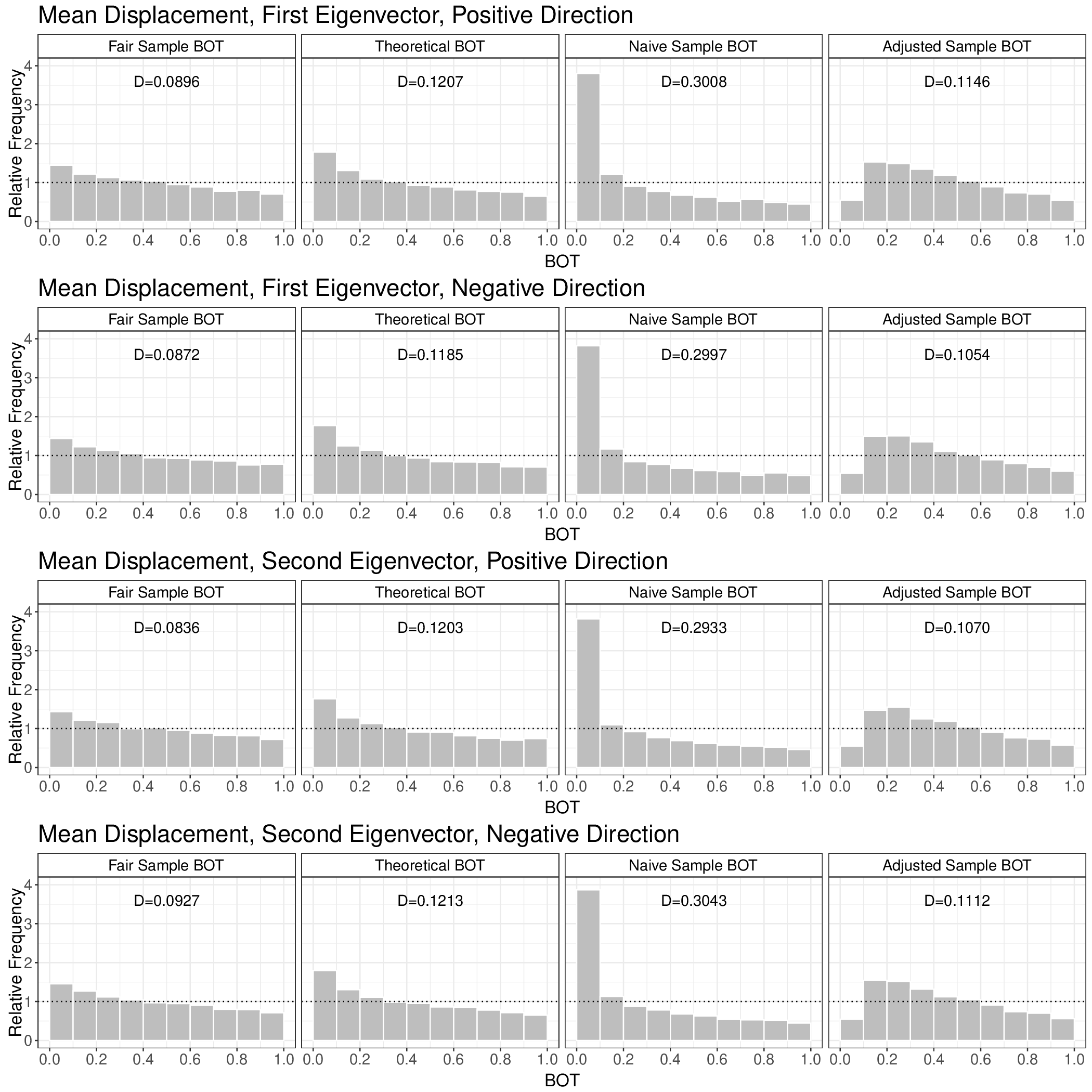, width=0.98\textwidth} 
   \caption{Histograms of various BOT versions for biased 3-dimensional 10-member ensemble forecasts together with the test statistic \ $D$ \ of the Kolmogorov-Smirnov test for uniformity.}
   \label{fig:BOTmeand3n10}
 \end{figure}
 
 Now we assume that the forecast covariances are correctly specified \ ($\boldsymbol\Sigma_f=\boldsymbol\Sigma_0$). \ Following \citet{wilks.2017}, we place the forecast vector mean \ $\boldsymbol\mu$ \  on the 15\,\% probability ellipsoid of the truth distribution in the direction of its first two principal axes. This means that
 \begin{equation*}
   \label{eq:mu}
   \boldsymbol\mu = \pm \sqrt{Q^{{\mathcal X}^2}_p(0.15)}\sqrt{\lambda_i}\boldsymbol e_i, \qquad i=1,2,
 \end{equation*}
 where \ $Q^{{\mathcal X}^2}_p(\alpha$) \ is the \ $\alpha$ \ quantile of a chi-square distribution with \ $p$ \ degrees of freedom and \ $\lambda_i$ \ and \ $\boldsymbol e_i$ \ denote the $i$th eigenvalue and eigenvector of \ $\boldsymbol\Sigma_0$, \ respectively \citep[see e.g.][Section 12.1]{wilks.2019}.

  There are no substantial differences between the various BOT versions when the  forecasts are biased and the ensemble size \ $n$ \ is much larger than the dimension \ $p$. \ For \ $n=50$ \ and \ $p=3$ \ (see Figure \ref{fig:BOTmeand3n50} in Section \ref{subsA.2} of the Appendix), only the naive sample BOT results in slightly larger Kolmogorov-Smirnov test statistic values than the other three variants. With the reduction of the ensemble size, similar to the case of a misspecified covariance structure (see Figure \ref{fig:BOTd3n10}), the situation changes. According to Figure \ref{fig:BOTmeand3n10}, while the fair sample BOT histograms do not change much and follow the shapes of the corresponding theoretical BOT histograms, the naive sample BOT starts putting most of the weight on the lower bins, whereas the adjusted sample BOT again underpopulates the lower bins and becomes hump-shaped.

 \section{Verification using operational ensemble forecasts}
 \label{sec4}

The properties of the naive, adjusted, and fair sample BOT are also investigated using various multivariate scenarios based on operational ECMWF ensemble forecasts. 

 \subsection{Ensemble forecasts and analyses}
 \label{subs4.1}
Similar to \citet{lb25}, we consider operational 100-member extended-range predictions of various weather quantities provided by the Integrated Forecast System \citep{ecmwf24} of the ECMWF at TCo319 horizontal resolution ($\approx$ 36 km) for the northern midlatitudes (35\degree\,N -- 65\degree\,N) for the period 1 September -- 30 November 2023. The forecasts are initialized at 0000 UTC and have a 46-day forecast range. Here, we focus  on lead times 1, 5, and 10 days. Based on their generation mechanism, the ensemble members can be considered statistically indistinguishable and thus exchangeable, sampling from the same predictive distribution.

We investigate BOT values for four different $p$-dimensional forecasts based on two  weather quantities with \ $p=2,4,9$, and $12$. \ To ensure a perfectly reliable situation, we consider a single ensemble member as the verification, and an $n$-member subsample of the remaining 99 members \ ($n\in \{8,16,24,32,64\}$) \ as the corresponding ensemble prediction. For forecasts sampled from a multivariate Gaussian law, such a setup should result in BOT values following a standard uniform distribution. Furthermore, we also study the shapes of the fair sample BOT histograms when the ensemble forecasts are verified against operational ECMWF analyses. 

First, we consider the two-dimensional vector wind on the 200 hPa level for each grid point of the studied ensemble domain. This configuration is referred to as uv200L2. 

The next two configurations are based on temperature forecasts at the 850 hPa level. To each point of the regular latitude-longitude grid, we associate a $2\times 2$ and a $3\times 3$ hypergrid (called a stencil), where the neigbouring points are obtained by approximately 1000 km meridional and longitudinal shifts. The predictions for locations in these stencils are then arranged in four- and nine-dimensional vectors, respectively, and the BOT values are calculated for these multivariate forecasts. The corresponding configurations are denoted as t850dxdyL4 and t850dxdyL9.

Finally, we also study profiles of vector wind forecasts on the 200, 300, 500, 700, 850, and 925 hPa pressure levels. We refer to this 12-dimensional configuration as uv200to925L12.

Note that the above four layouts are among the thirteen configurations studied in \citet{lb25}, where the multivariate normality of the corresponding forecast vectors is also assessed using the Henze–Zirkler (HZ) test \citep{henze.zirkler.1990}.

\begin{figure}[!th]
 \centering
   \epsfig{file=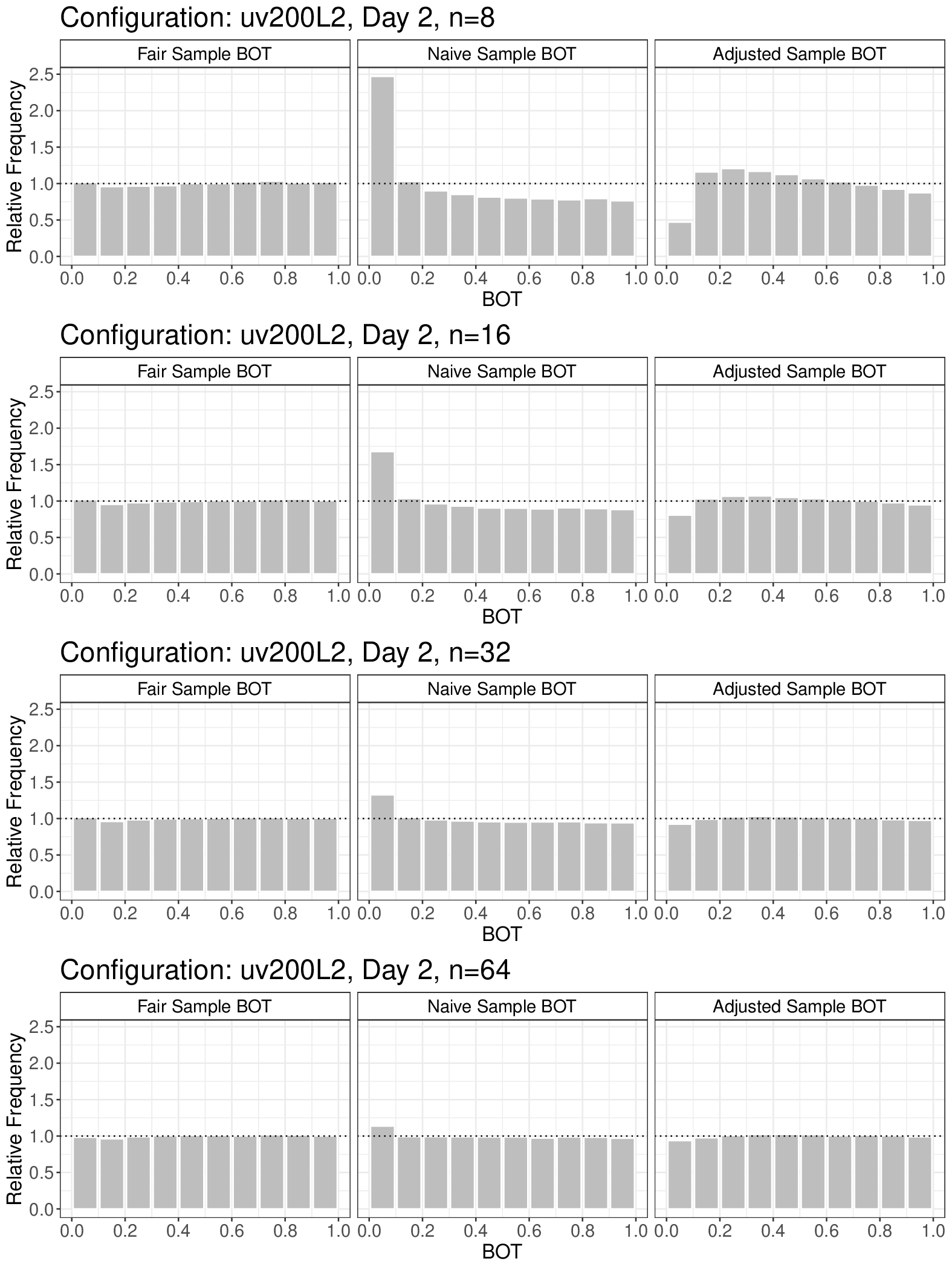, width=.735\textwidth} 
   \caption{Histograms of various BOT versions for the two-dimensional predictand of vector wind at the 200 hPa level (configuration uv200L2) for day 2.}
   \label{fig:BOTuv200L2D2}
 \end{figure}
 
 \subsection{Results}
\label{subs4.2}
 
In this section, we compare the shapes of the histograms of the fair sample BOT values \eqref{eq:fairBOT} for perfectly reliable ensemble forecasts with the forms of the corresponding naive and adjusted sample BOT histograms for the configurations introduced in Section \ref{subs4.1}. Furthermore, we provide results for the fair sample BOT when forecasts are verified against analyses.

\subsubsection{Vector wind}
\label{subs4.2.1}

Figure \ref{fig:BOTuv200L2D2} displays histograms of the three investigated sample BOT versions corresponding to two-day ahead two-dimensional horizontal vector wind ensemble forecasts comprising \ $n=8, \ 16, \ 32$, and $64$ members. According to the results of the HZ test reported in \citet[][Section 4.5]{lb25}, this weather quantity can be considered bivariate Gaussian. The histogram of the fair sample BOT is almost perfectly flat for all investigated ensemble sizes, whereas the shapes of the naive- and adjusted sample BOT histograms align with the corresponding plots in the top panels of Figures \ref{fig:BOTd3n50} -- \ref{fig:BOTd30n50}. For 8-member forecasts, the naive sample BOT values exhibit a rather strong bias in the direction of the lower bins, which bias is gradually reduced as the ratio of the ensemble size \ $n$ \ and dimension \ $p$ \ increases. In contrast to its naive counterpart, for the smallest ensemble size, the adjusted sample BOT histogram is hump-shaped with an underpopulated lowest bin. Again, with the increase of the ensemble size, this deficiency considerably improves. Nevertheless, not even 64-member forecasts can provide flat BOT histograms; slight traces of the hump shape still remain. Similar behaviour can be observed on BOT histograms corresponding to day 5 and day 10; see Figures \ref{fig:BOTuv200L2D5} and \ref{fig:BOTuv200L2D10} in Section \ref{subsB.1} of the Appendix, respectively.

\subsubsection{Stencils}
\label{subs4.2.2}

\begin{figure}[!th]
 \centering
   \epsfig{file=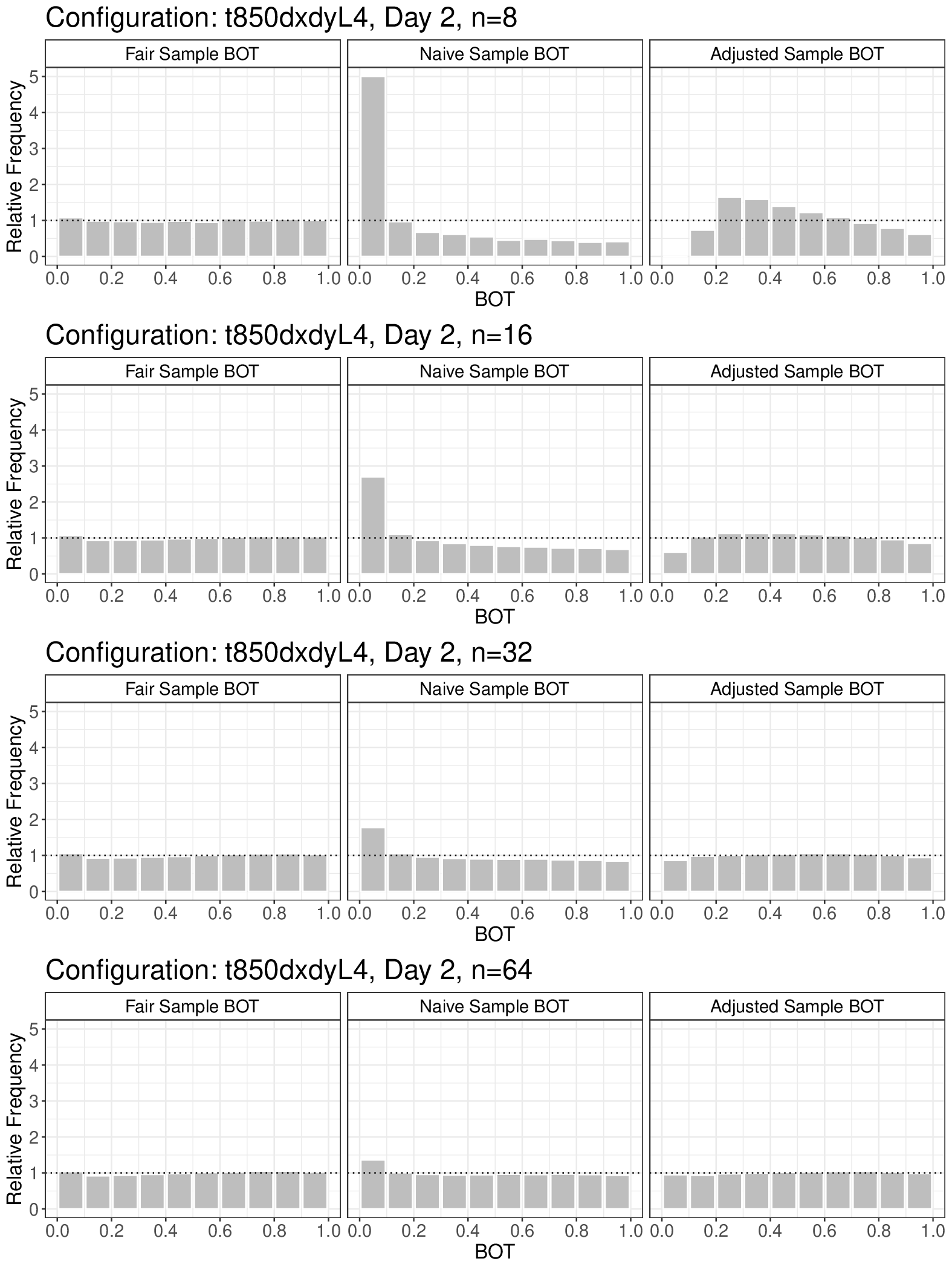, width=.735\textwidth} 
   \caption{Histograms of various BOT versions for the four-dimensional predictand of 850 hPa temperature on the $2\times 2$ stencil of points (configuration t850dxdyL4) for day 2.}
   \label{fig:BOTt850dxdyL4D2}
 \end{figure}

 \begin{figure}[!th]
 \centering
   \epsfig{file=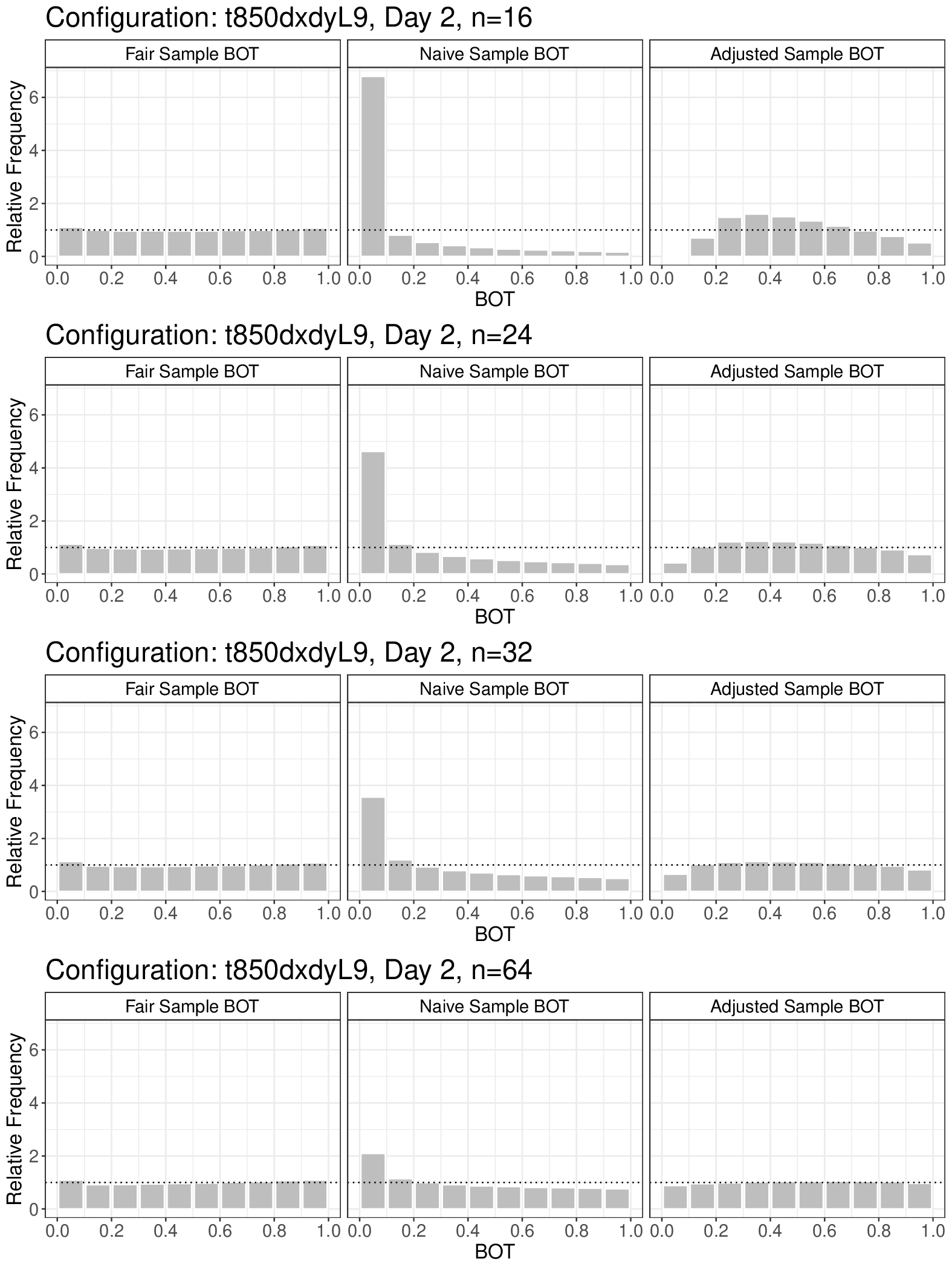, width=.735\textwidth} 
   \caption{Histograms of various BOT versions for the nine-dimensional predictand of 850 hPa temperature on the $3\times 3$ stencil of points (configuration t850dxdyL9) for day 2.}
   \label{fig:BOTt850dxdyL9D2}
 \end{figure}

The two-day ahead four- and nine-dimensional forecasts of 850 hPa temperature on the $2\times 2$ and $3\times 3$ stencils of points leading to the BOT histograms depicted in Figures \ref{fig:BOTt850dxdyL4D2} and \ref{fig:BOTt850dxdyL9D2}, respectively, can still be considered Gaussian \citep[see again,][Section 4.5]{lb25}. In the four-dimensional case (Figure \ref{fig:BOTt850dxdyL4D2}), the shape of the fair sample BOT histogram does not depend on the ensemble size and exhibits just a very minor deviation from uniformity. Now, for the smallest ensemble size of \ $n=8$, \ the lower ratio of size to dimension results in an even more biased naive sample BOT histogram than in Figure \ref{fig:BOTuv200L2D2}. This ratio as well as the ensemble size appear to determine the deviation from uniformity for the naive sample BOT. The same phenomenon can also be observed for the adjusted sample BOT values. For the eight-member ensemble, the hump shape is very pronounced, and the lowest bin is almost empty, whereas for ensemble sizes $16$, $32$, and $64$, the histograms are very similar to the adjusted sample BOT histograms of Figure \ref{fig:BOTuv200L2D2} corresponding to \ $n=8, \ 16$, and $32$, respectively.

\begin{figure}[!th]
 \centering
   \epsfig{file=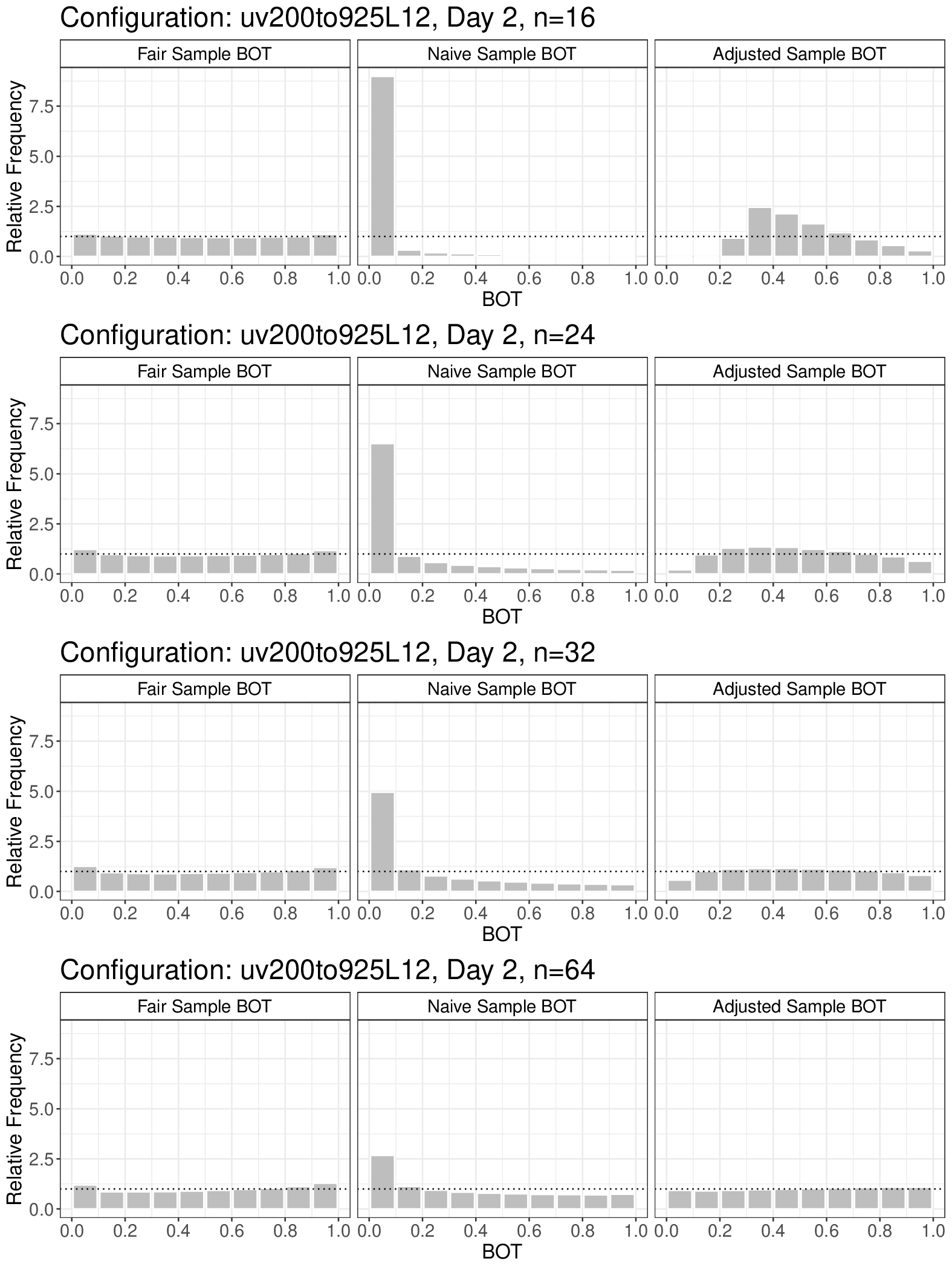, width=.735\textwidth} 
   \caption{Histograms of various BOT versions for the 12-dimensional predictand of horizontal vector wind on the levels of 200, 300, 500, 700, 850, and 925 hPa (configuration uv200to925L12) for day 2.}
   \label{fig:BOTuv200to925L12D2}
 \end{figure}
 
According to Figure \ref{fig:BOTt850dxdyL9D2}, $16$-, $24$-, $32$-, and $64$-member nine-dimensional ensemble forecasts of 850 hPa temperature again result in identical fair sample BOT histograms, which are now very slightly $\cup$-shaped. Despite the extreme bias towards the lower bins of the naive sample BOT corresponding to the $16$-member forecasts considerably reduces with the increase of the ensemble size, it is still rather pronounced even for the largest ensemble. However, one should note that the size/dimension ratio of $64/9$ of the latter is worse than the same ratio of the $32$-member four-dimensional forecast of 850 hPa temperature on the $2\times 2$ stencil (see Figure \ref{fig:BOTt850dxdyL4D2}). Finally, for the smallest sample size the adjusted sample BOT histogram is again strongly asymetric and hump shaped with highly underpopulated lower bins; nevertheless, these deficiencies almost completely fade as the ensemble size increases.
 
\begin{figure}[!th]
 \centering
   \epsfig{file=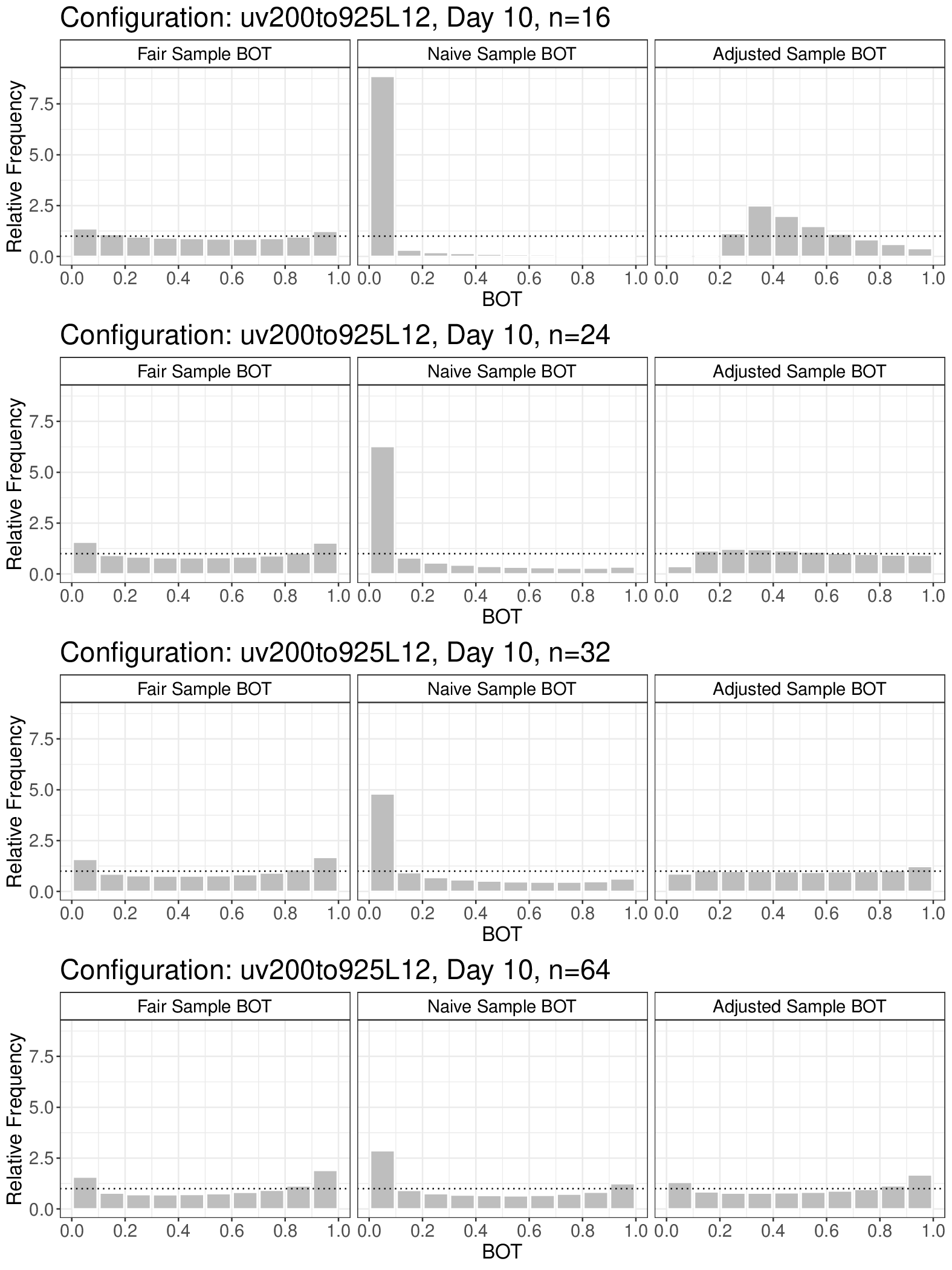, width=.735\textwidth} 
   \caption{Histograms of various BOT versions for the 12-dimensional predictand of horizontal vector wind on the levels of 200, 300, 500, 700, 850, and 925 hPa (configuration uv200to925L12) for day 10.}
   \label{fig:BOTuv200to925L12D10}
 \end{figure}

 The BOT histograms of five- and ten-day ahead forecasts of 850 hPa temperature on the $2\times 2$ and $3\times 3$ stencils of points convey the same message; see Section \ref{subsB.2} of the Appendix.

 \subsubsection{Vector wind profiles}
 \label{subs4.2.3}

\begin{figure}[!th]
 \centering
   \epsfig{file=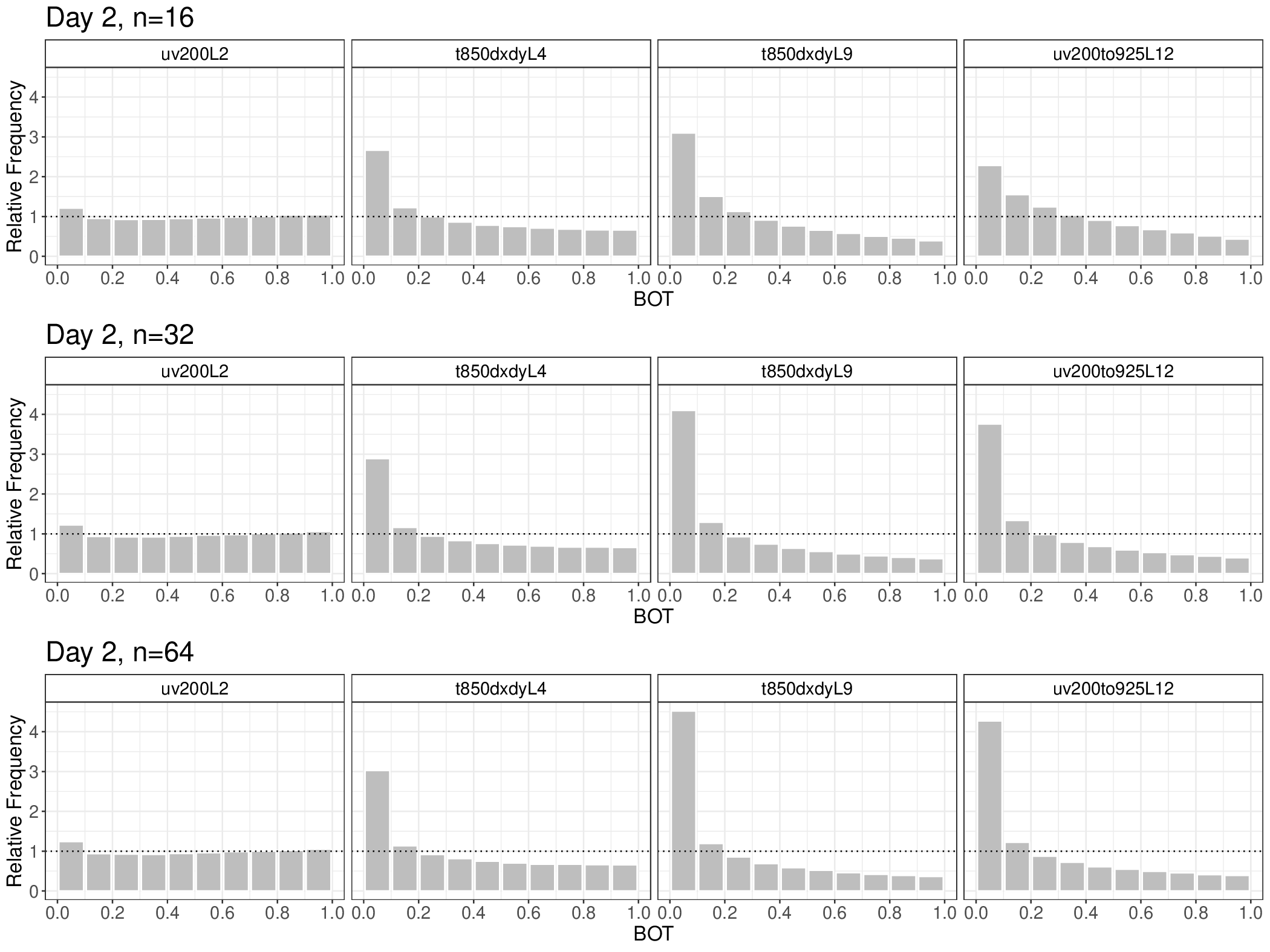, width=.98\textwidth} 
   \caption{Histograms of the fair sample BOT for various multivariate predictands verified against analyses for day 2.}
   \label{fig:fairBOTallD2}
 \end{figure}

The fair sample BOT histograms of the 12-dimensional forecasts of horizontal vector wind on the levels of 200, 300, 500, 700, 850, and 925 hPa for day 2 depicted in the left column of Figure \ref{fig:BOTuv200to925L12D2} are almost identical for all considered ensemble sizes \ ($n=16, \ 24, \ 32, \ 64$) \ exhibiting a very slight $\cup$-shape. According to Figure 6 of \citet{lb25}, the deviation of this prediction from multivariate normality is higher than for the previous examples and almost significant at a 5\,\% level. For the smallest ensemble size with a very low size to dimension ratio of 4/3, nearly all naive sample BOT values are concentrated in the lowest two bins. In contrast, for \ $n=16$, \ the adjusted sample BOT values avoid these two bins, and the hump shape and bias of the histogram are more pronounced than for the previously studied configurations. 
The bias of the naive sample BOT remains rather strong even for \ $n=64$, \ whereas for the adjusted sample BOT, the deviation from uniformity almost completely disappears as the ensemble size increases.

The BOT histograms of Figure \ref{fig:BOTuv200to925L12D10} corresponding to 10-day ahead forecasts of the same 12-dimensional configuration tell a slightly different story. All fair sample BOT histograms are $\cup$-shaped, which deficiency is getting more pronounced with the increase of the sample size. The left extreme bias of the naive and the strong hump shape of the adjusted sample BOT for \ $n=16$ \ also transform to $\cup$-shapes as the ensemble size reaches its largest value of $64$, with moderate left and slight bias, respectively. The deviation from the uniformity of all fair sample BOT histograms and the adjusted BOT histogram for the largest ensemble size is expected to originate from the highly significant deviation of the forecast at day 10 from the multivariate normality, confirmed by the HZ test \citep[see again][Figure 6]{lb25}.

Note that the distribution of the five-day ahead forecasts of horizontal vector wind on the levels of 200, 300, 500, 700, 850, and 925 hPa still cannot be considered Gaussian; however, the deviation is smaller than at day 10. Accordingly, the corresponding BOT histograms (Figure \ref{fig:BOTuv200to925L12D5} in Section \ref{subsB.3} of the Appendix) show an intermediate state; the fair sample BOT histograms and the adjusted sample BOT histogram for the 64-member ensemble are still $\cup$-shaped, but the divergence from flatness is not as strong as for the longer lead time.

\subsubsection{Verification against analyses}
\label{subs4.2.4}

Figure \ref{fig:fairBOTallD2} displays the fair sample BOT histograms of two-day ahead 16-, 32-, and 64-member ensemble forecasts of all four investigated configurations, now verified against analyses. In the case of the vector wind  on the 200 hPa level (configuration uv200L2), the shape of the histogram does not depend on the ensemble size and is almost flat, showing just a very mild $\cup$-shape. This suggests that the forecasts are nearly calibrated and sample a Gaussian distribution close to the one of the verifying analyses. 

In contrast, the other three predictands (t850dxdyL4, t850dxdyL9, uv200to925L12) exhibit positively skewed histograms peaking at low values. Further examination of the data shows evidence of larger univariate miscalibration for these predictands than for vector wind at 200 hPa. The magnitude of the bias normalised by the standard deviation of the ensemble is approximately twice as large than for uv200L2. The shape of the histograms  is consistent with that in the simulation studies that consider a mean displacement of the distribution (Fig.~\ref{fig:BOTmeand3n10}). We have also compared the mean ensemble covariances with the second moments of the error of the ensemble mean and found no indications of substantial miscalibration of the predicted covariances at Day~2 for the four configurations shown here.

\section{Discussion and conclusions}
\label{sec5}

Compared to the univariate situation, one has to approach the assessment of calibration of multivariate probabilistic more carefully, as there are more complex forms of miscalibration, such as bias in the mean vector, misspecified covariance structure, or the combination of these deficiencies. In the present work, we focus on ensemble predictions following a \ $p$-dimensional multivariate Gaussian law and the Box ordinate transform based on the Mahalanobis distance of the ensemble mean and the corresponding verifying observation vector. For reliable forecasts, sampled from the same Gaussian law as the observation, the asymptotic distribution of the Mahalanobis distance is a chi-square with \ $p$ \ degrees of freedom, so for large ensemble sizes \ $n$ \ (\citet{wilks.2017} suggests having \ $n\approx 100$ \ for \ $p=3$), \ the sample BOT relying on the estimated mean vector and covariance matrix, is asymptotically standard uniform. 

Here, we derived the exact distribution of the Gaussian Mahalanobis distance and, based on this, suggested an ensemble size-dependent fair version of the sample BOT. Hence, a reliable forecast of the multivariate normal distribution will result in a flat histogram of fair BOT values regardless of ensemble size. This has been confirmed by our simulation study involving various combinations of the ensemble size and forecast dimension. Results have also shown that the shape of the fair sample BOT histograms is qualitatively similar to that of the theoretical BOT histograms of the distribution for several unreliable scenarios, which is not the case for the classical sample BOT versions for ratios of ensemble size \ $n$ \ and dimension \ $p$. However, the magnitude of the departures from the flat histogram varies with ensemble size.

Furthermore, while reliability for multivariate forecasts will always imply flat fair sample BOT histograms, the converse is not true. There are cases where the forecast is not reliable, and yet the histogram indicates a standard uniform law. Such situations can arise from compensating errors that are masked by the projection into a single rank histogram (in our simulation study, it appears when either marginal variances or correlations are loaded with additive errors having alternating signs across the dimensions), which is analogous to compensating errors affecting univariate rank histograms \citep{hamill.2001}. Thus, we recommend combining the fair BOT histogram with (i) univariate rank histograms for marginal distributions and (ii) fair BOT histograms for lower-dimensional subspaces. This will reduce the likelihood of falsely concluding reliability and also helps to better understand the causes of deviations from reliability.

The performance of the fair sample BOT has also been tested using ECMWF IFS extended-range forecasts of 850 hPa temperature and horizontal vector wind at six different pressure levels by forming  2-, 4-, 9-, and 12-dimensional vector predictands. Similar to \citet{lb25}, the multivariate normality of these forecasts was verified using the Henze-Zirkler test, where the test statistic was also applied to quantify the deviation from the desired Gaussian law. We have considered the perfectly reliable situation when one of the ensemble members plays the role of the observation. These experiments have revealed that while the fair sample BOT exhibits some robustness against deviations from multivariate normality when the ensemble forecast is perfectly reliable, considerable deviations clearly result in departures from flat fair sample BOT histograms. In these situations, probably more general approaches would be needed, like, for instance, 2D rank histograms proposed by \cite{benbouallegue2025}.

Finally, when verified against analyses, the shapes of the fair sample BOT histograms correctly reflect the situation when the forecast is nearly calibrated and samples a Gaussian law, and the situation of a miscalibrated forecast due to the presence of bias.

Although the present results are valid for forecasts following a multivariate Gaussian law, one might try to extend them for non-Gaussian setups. A possible approach is to apply a Box-Cox transformation to the marginals to achieve approximate normality, this method proved to be successful in hydrological forecasting; see, for instance, \citet{dags07,bhea19}, or for a multivariate setup, \citet{hfz13}.

The introduced fair BOT can also be utilized to verify post-processed multivariate forecasts. Here, one can think of either the classical two-step approaches, when after univariate post-processing of the marginals, the lost dependencies are restored by applying empirical or parametric copulas on samples from the calibrated one-dimensional predictive distributions \citep{lerch.ea.2020,llhb.2023} or of the state-of-the-art machine-learning-based generative models \citep[see e.g.][]{dh21,cjsl24}.

\section*{Acknowledgements}
S\'andor Baran was supported by the Hungarian National Research, Development and Innovation Office under Grant K142849. He is also grateful to the ECMWF for supporting his research stay in Reading, United Kingdom. 

 \section*{Data availability statement}
 The data used in this study consists of archived operational data from ECMWF. It is available under a \href{https://creativecommons.org/licenses/by/4.0/}{CC BY 4.0 license} and access can be requested via ECMWF's web archive (\url{https://apps.ecmwf.int/archive-catalogue/?class=od}).



\bibliographystyle{rss}

\begin{figure}[!h]
 \centering
   \epsfig{file=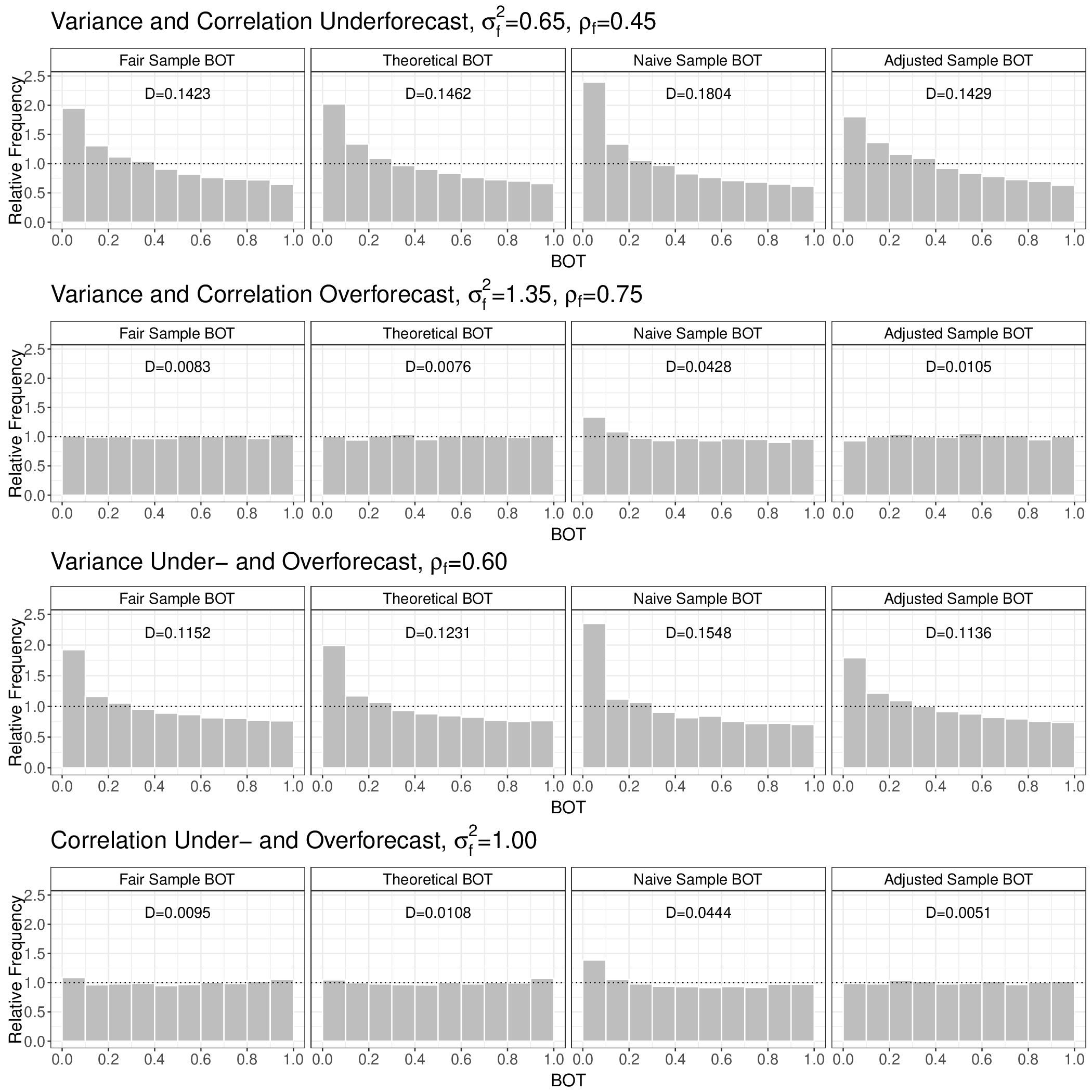, width=.98\textwidth} 
   \caption{Histograms of various BOT versions for misspecified 3-dimensional 50-member ensemble forecasts together with the test statistic \ $D$ \ of the Kolmogorov-Smirnov test for uniformity.}
   \label{fig:BOTmixd3n50}
 \end{figure}

 \begin{figure}[!h]
 \centering
   \epsfig{file=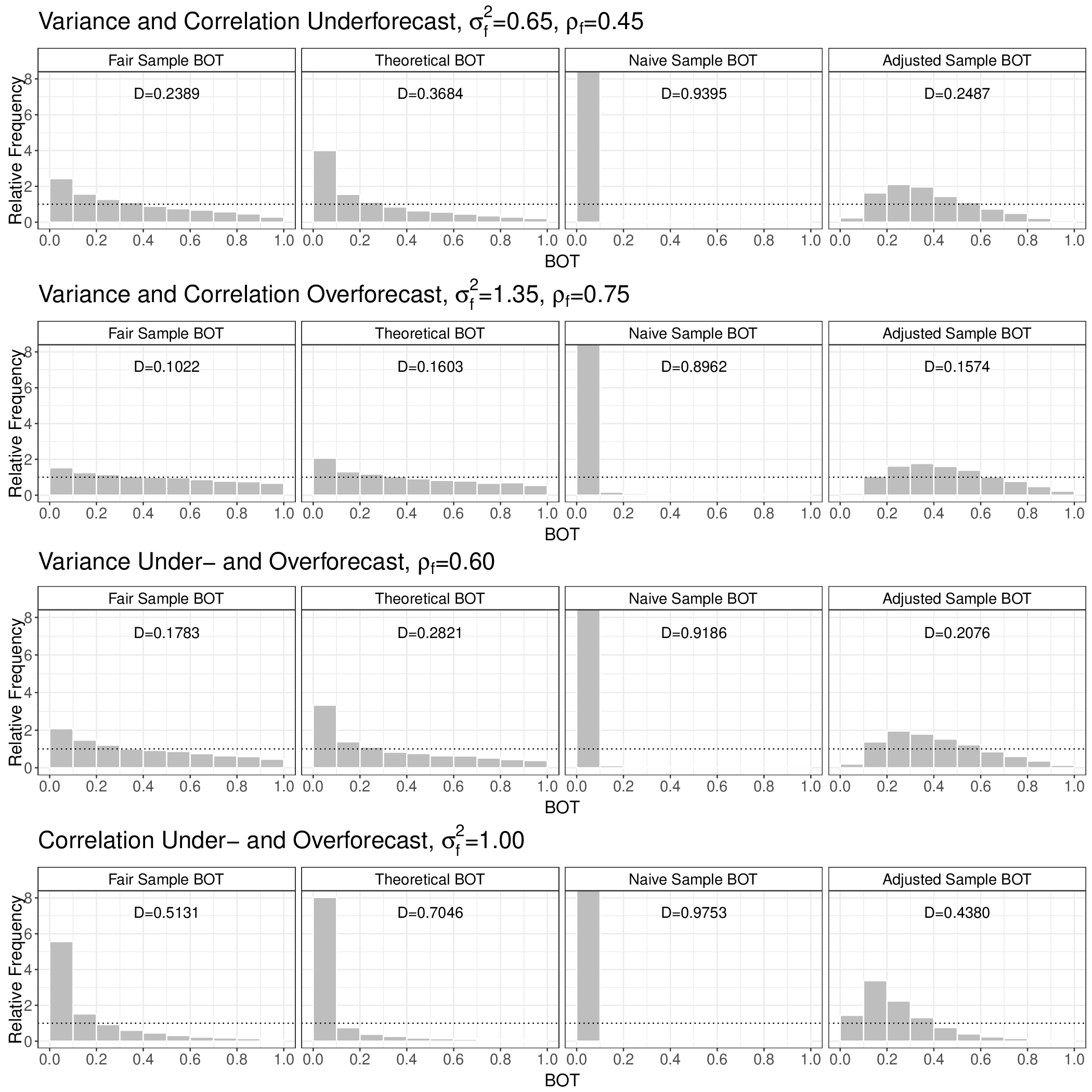, width=.98\textwidth} 
   \caption{Histograms of various BOT versions for misspecified 30-dimensional 50-member ensemble forecasts together with the test statistic \ $D$ \ of the Kolmogorov-Smirnov test for uniformity.}
   \label{fig:BOTmixd30n50}
 \end{figure}

 \begin{figure}[!h]
   \centering
   \epsfig{file=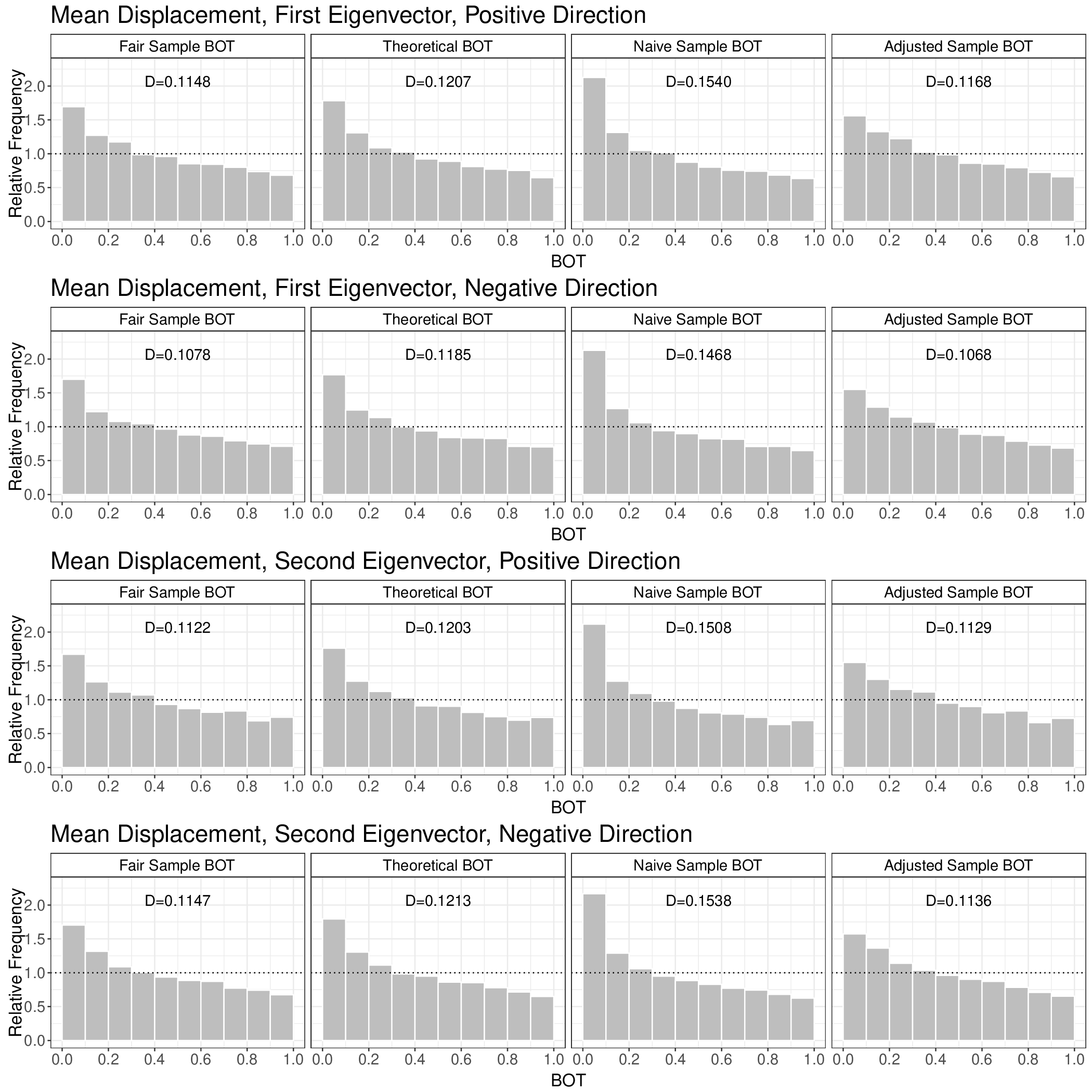, width=0.98\textwidth} 
   \caption{Histograms of various BOT versions for biased 3-dimensional 50-member ensemble forecasts together with the test statistic \ $D$ \ of the Kolmogorov-Smirnov test for uniformity.}
   \label{fig:BOTmeand3n50}
 \end{figure}

 \begin{figure}[!h]
 \centering
   \epsfig{file=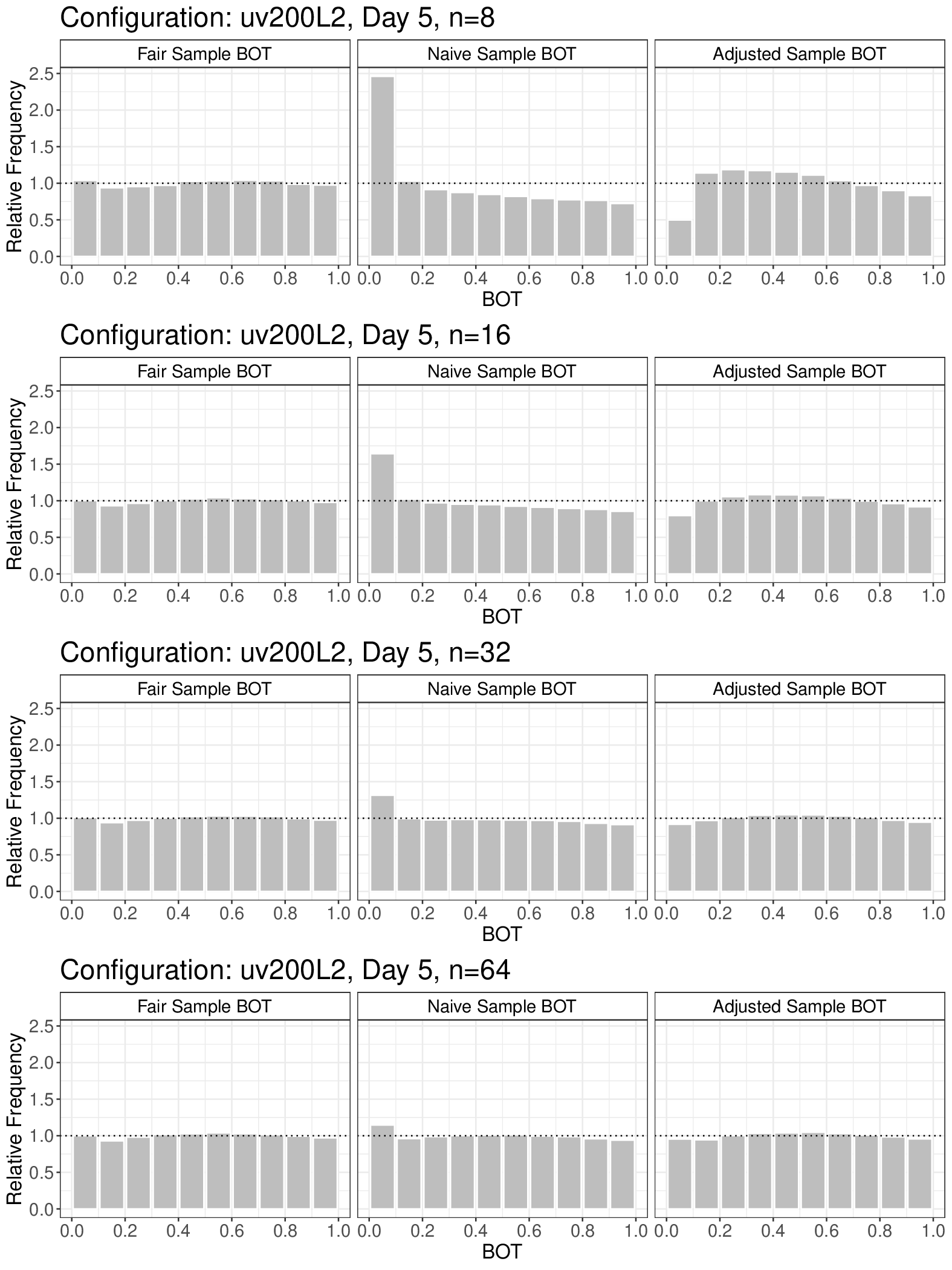, width=.735\textwidth} 
   \caption{Histograms of various BOT versions for the two-dimensional predictand of vector wind at the 200 hPa level (configuration uv200L2) for day 5.}
   \label{fig:BOTuv200L2D5}
 \end{figure}

 \begin{figure}[!h]
 \centering
   \epsfig{file=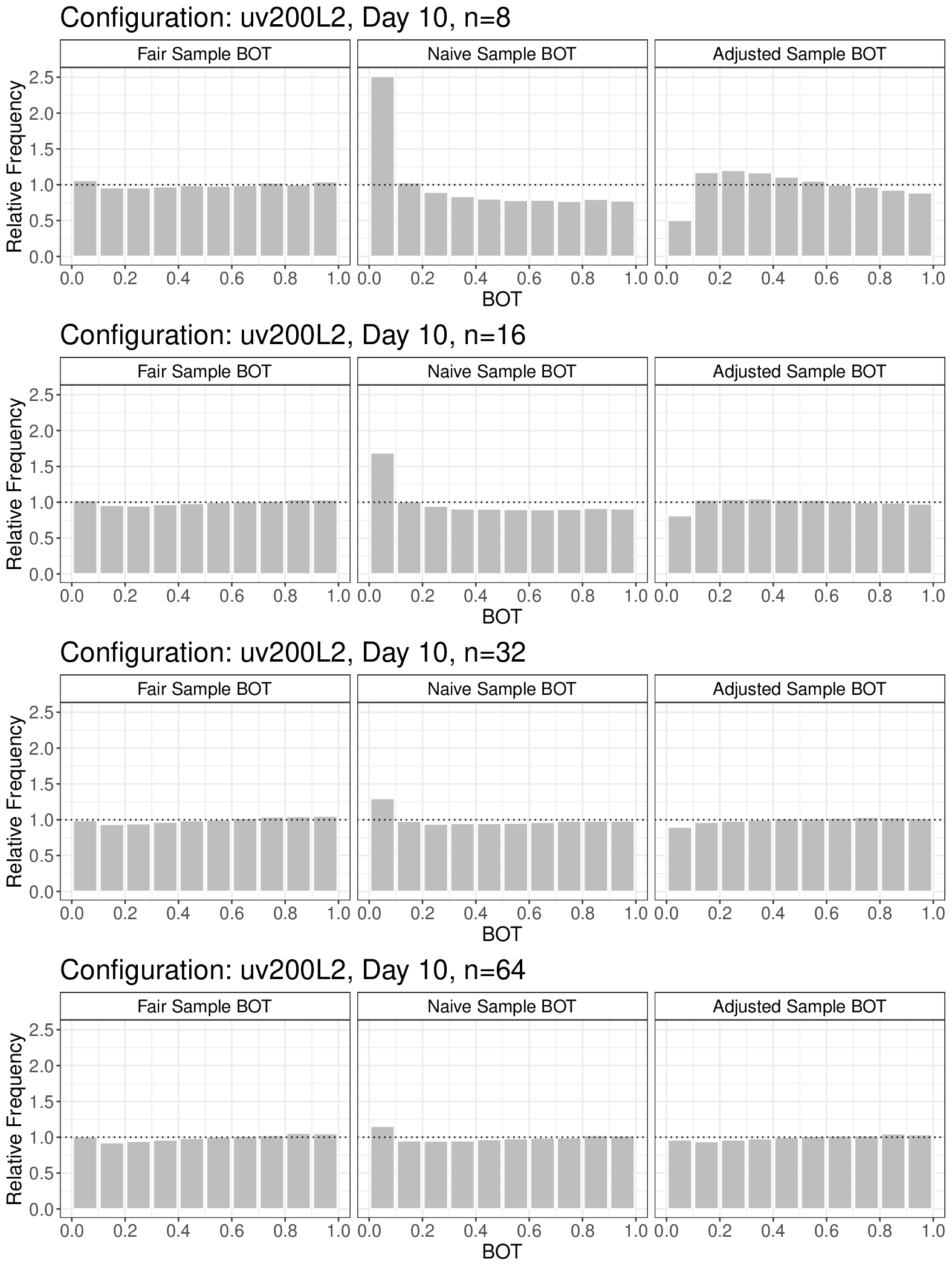, width=.735\textwidth} 
   \caption{Histograms of various BOT versions for the two-dimensional predictand of vector wind at the 200 hPa level (configuration uv200L2) for day 10.}
   \label{fig:BOTuv200L2D10}
 \end{figure}

 \begin{figure}[!h]
 \centering
   \epsfig{file=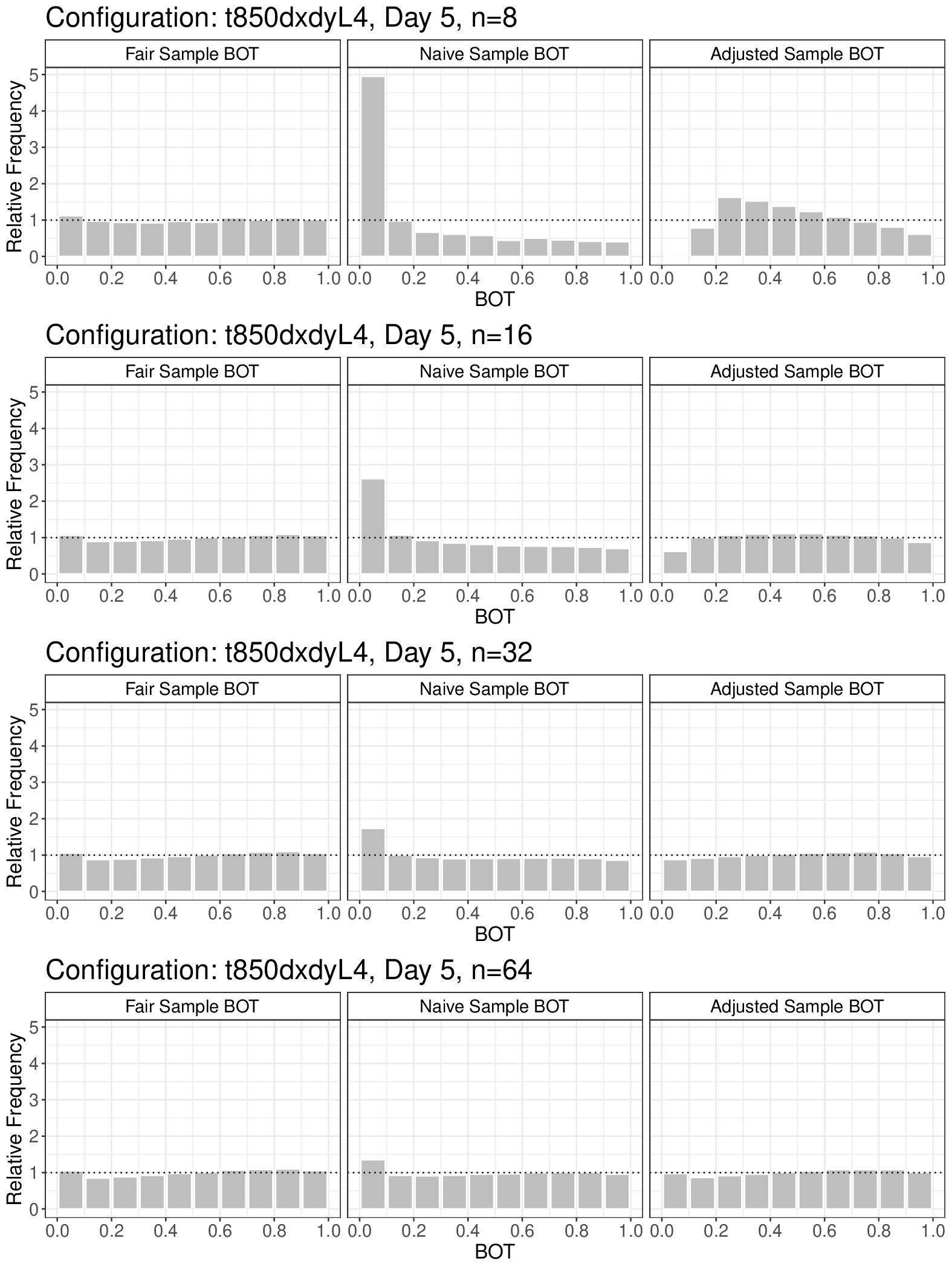, width=.735\textwidth} 
   \caption{Histograms of various BOT versions for the four-dimensional predictand of 850 hPa temperature on the $2\times 2$ stencil of points (configuration t850dxdyL4) for day 5.}
   \label{fig:BOTt850dxdyL4D5}
 \end{figure}

 \begin{figure}[!h]
 \centering
   \epsfig{file=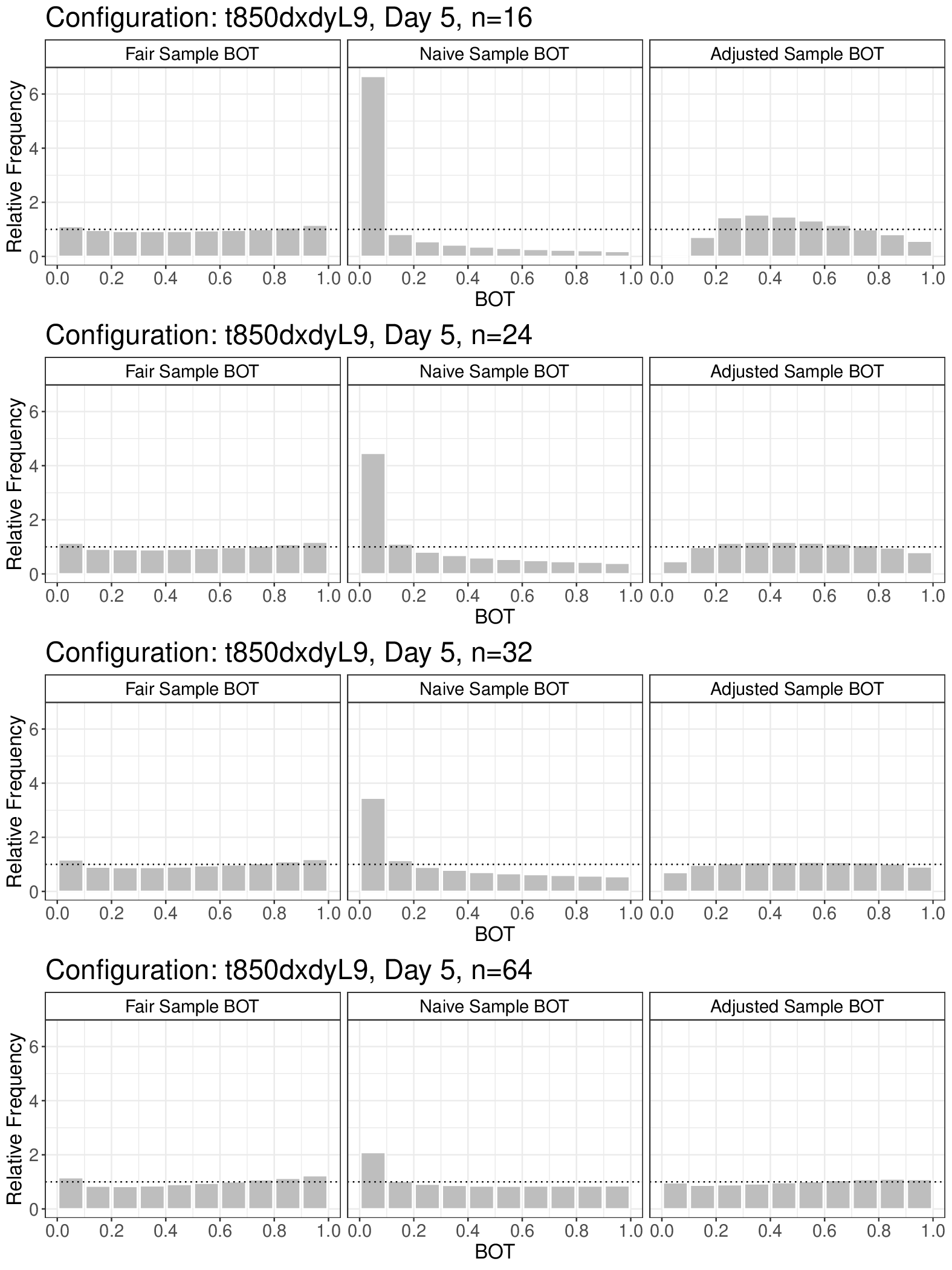, width=.735\textwidth} 
   \caption{Histograms of various BOT versions for the four-dimensional predictand of 850 hPa temperature on the $3\times 3$ stencil of points (configuration t850dxdyL9) for day 5.}
   \label{fig:BOTt850dxdyL9D5}
 \end{figure}
 
\bibliography{refsarXiv}

\begin{appendix}
 \section{Additional simulations}
  \label{secA}
Here we show BOT histograms of further simulation studies, part of which are discussed in Section \ref{subs3.2}.

\begin{figure}[!h]
 \centering
   \epsfig{file=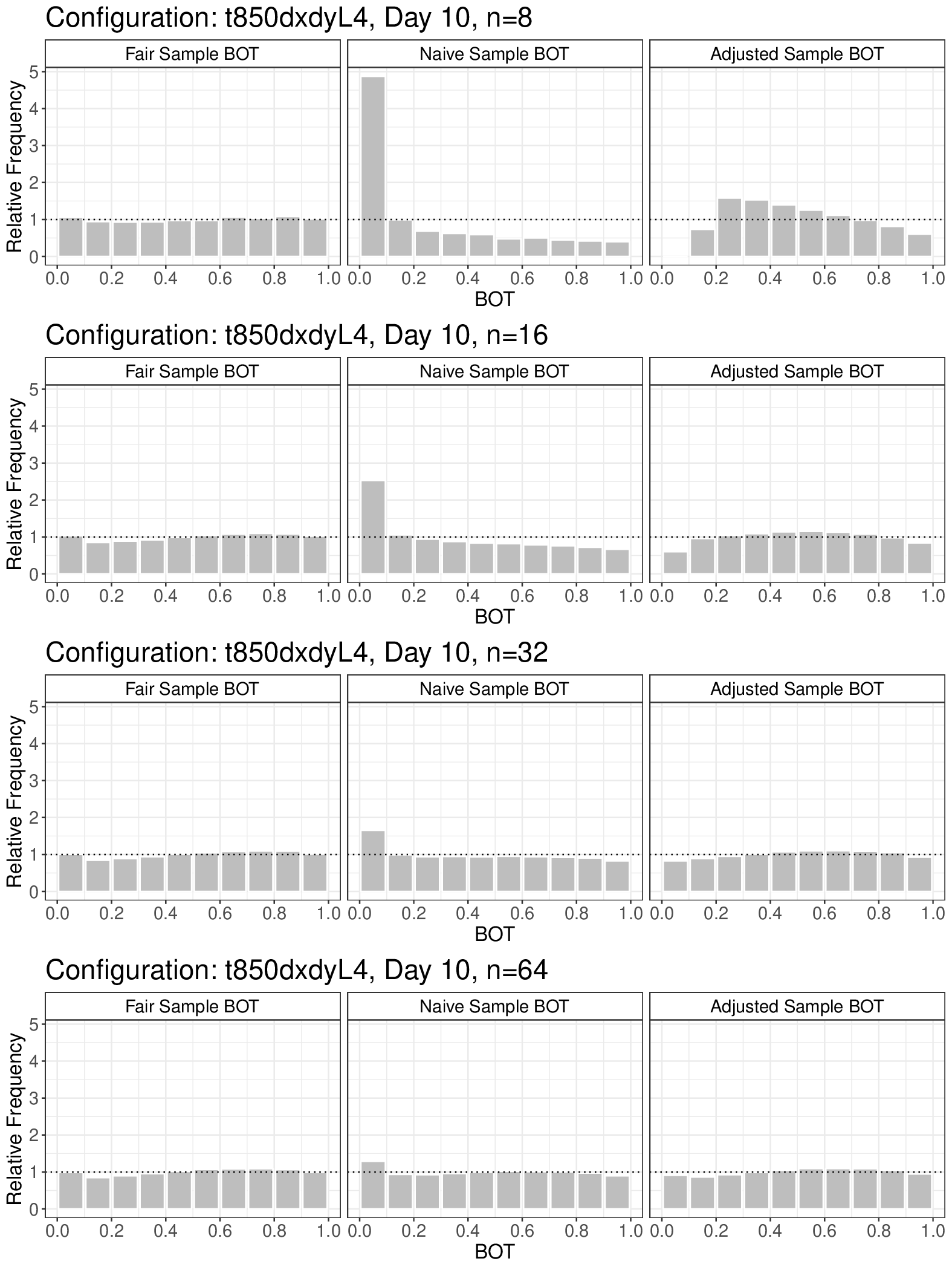, width=.735\textwidth} 
   \caption{Histograms of various BOT versions for the four-dimensional predictand of 850 hPa temperature on the $2\times 2$ stencil of points (configuration t850dxdyL4) for day 10.}
   \label{fig:BOTt850dxdyL4D10}
 \end{figure}

\subsection{Misspecified covariance structure}
 \label{subsA.1}

 \begin{figure}[!h]
 \centering
   \epsfig{file=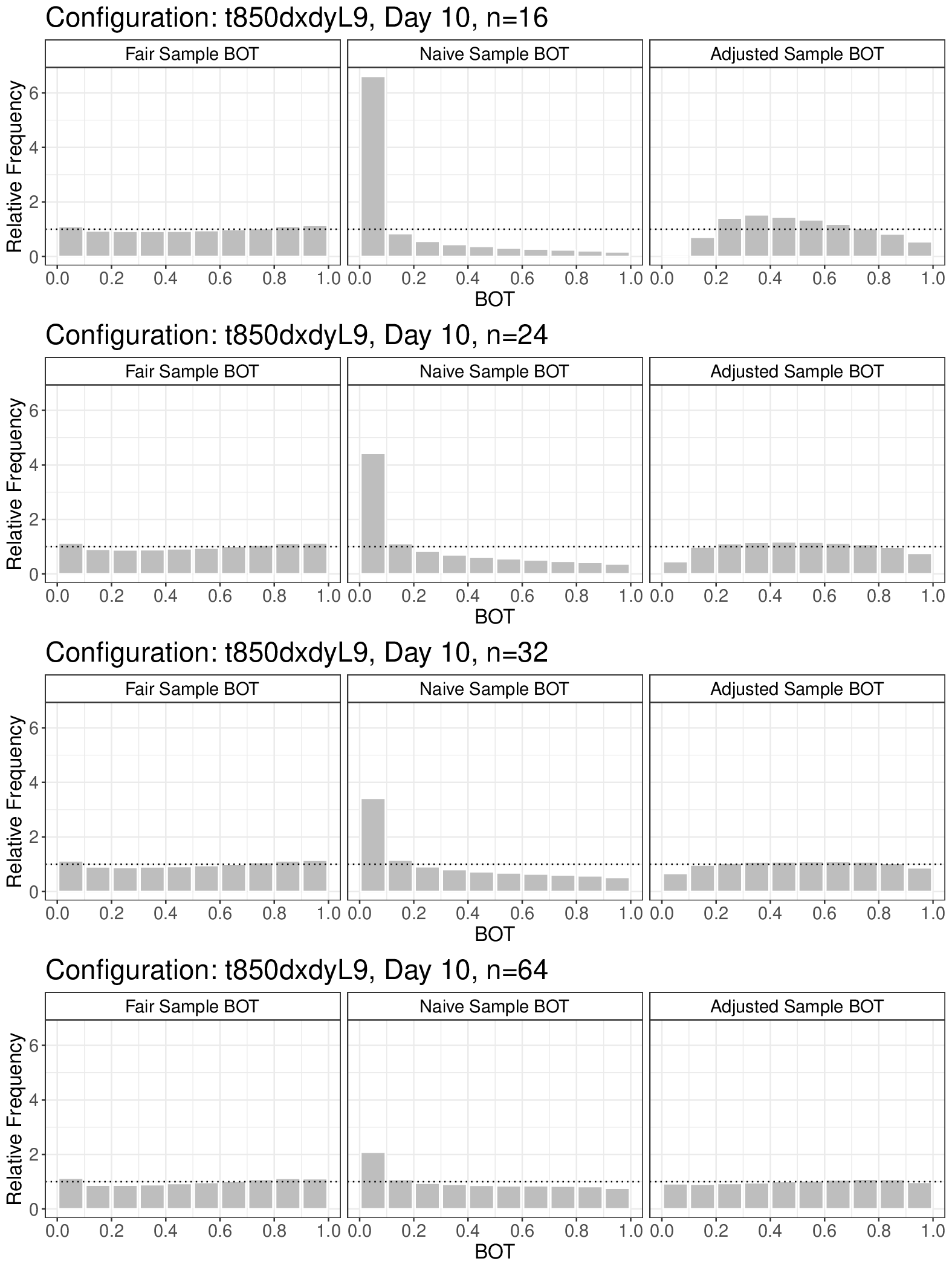, width=.735\textwidth} 
   \caption{Histograms of various BOT versions for the four-dimensional predictand of 850 hPa temperature on the $3\times 3$ stencil of points (configuration t850dxdyL9) for day 10.}
   \label{fig:BOTt850dxdyL9D10}
 \end{figure}
 
 As discussed in Section \ref{subs3.2.1}, for 3-dimensional 50-member ensemble forecasts, the histograms of the fair sample BOT, theoretical BOT, and adjusted sample BOT are consistent in each of the four miscalibration schemes depicted in Figure \ref{fig:BOTmixd3n50}, just the naive sample BOT displays slightly more overpopulated lower bins.

 A different situation can be observed in Figure \ref{fig:BOTmixd30n50}. The very low ensemble size to dimension ratio in each of the four investigated cases of miscalibration results in an extreme bias in the naive sample BOT values and strongly hump-shaped and biased adjusted sample BOT histograms. These shapes very much differ from the forms of corresponding histograms of the theoretical BOT values. The fair sample BOT does a better job of mimicking the theoretical one; however, the bias towards the lower bins of the histograms is less pronounced, resulting in consistently smaller values of Kolmogorov-Smirnov \ $D$ \ statistic. Finally, note that in contrast to the 3-dimensional cases with much better \ $n/p$ \ ratios, here none of the BOT versions lead to confusing flat histograms.

 \subsection{Bias}
   \label{subsA.2}
   
  \begin{figure}[!h]
 \centering
   \epsfig{file=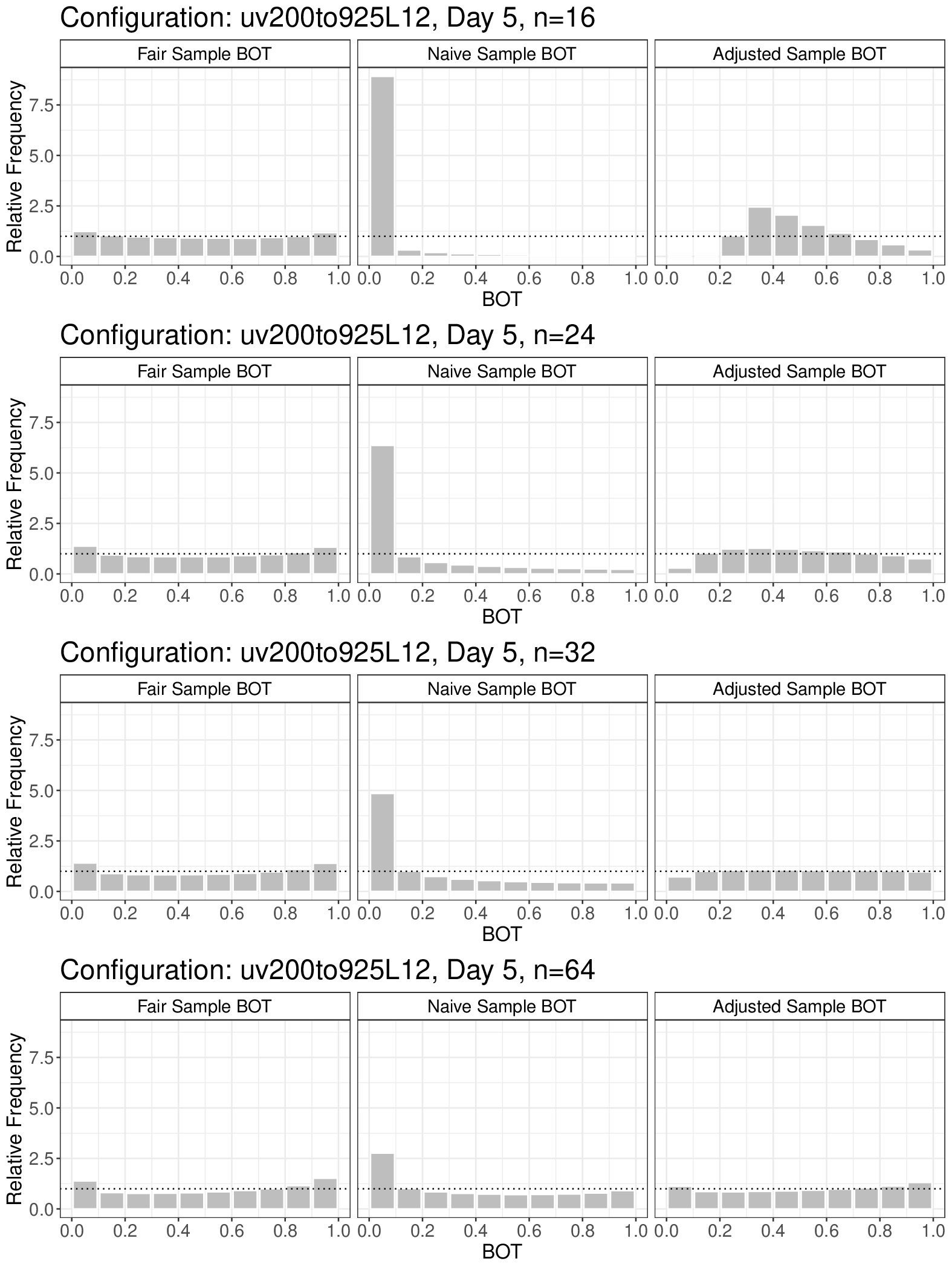, width=.735\textwidth} 
   \caption{Histograms of various BOT versions for the 12-dimensional predictand of horizontal vector wind on the levels of 200, 300, 500, 700, 850, and 925 hPa (configuration uv200to925L12) for day 5.}
   \label{fig:BOTuv200to925L12D5}
 \end{figure}

 As discussed in Section \ref{subs3.2.2}, Figure \ref{fig:BOTmeand3n50}, which provides BOT histograms for biased 3-dimensional 50-member ensemble forecasts tells the same story as the matching Figures \ref{fig:BOTd3n50} and \ref{fig:BOTmixd3n50} corresponding to various misspecifications of the covariance structure. In each of the four studied cases all BOT versions, but the naive sample BOT result in similar histograms, whereas the left bias of the naive sample BOT values is consistently stronger, leading to slightly larger values of the Kolmogorov-Smirnov test statistic.

 \begin{figure}[!th]
 \centering
   \epsfig{file=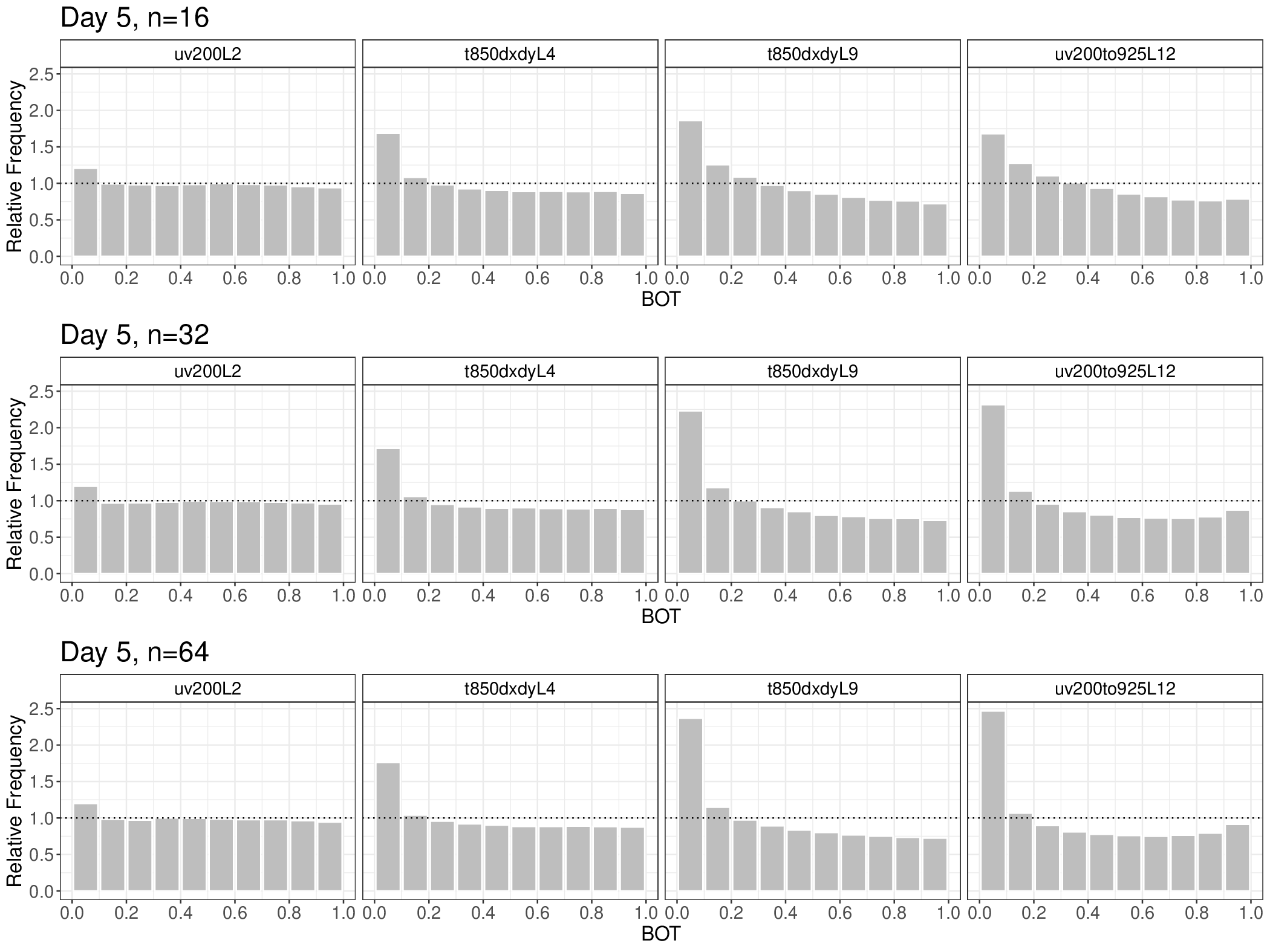, width=.98\textwidth} 
   \caption{Histograms of the fair sample BOT for various multivariate predictands verified against analyses for day 5.}
   \label{fig:fairBOTallD5}
 \end{figure}

  \section{BOT histograms for operational ensemble forecasts}
  \label{secB}

  Section \ref{subs4.2}, reporting the behaviour of various sample BOT versions for operational ECMWF ensemble forecasts, focuses mostly on two-day ahead predictions. Below we provide some additional results for longer lead times.

  \subsection{Vector wind}
  \label{subsB.1}

 As mentioned in Section \ref{subs4.2.1}, in the case of vector wind at the 200 hPa level, there are no substantial differences in the behaviour of the various BOT versions for the investigated lead times, Figures \ref{fig:BOTuv200L2D5} and \ref{fig:BOTuv200L2D10} are very similar to Figure \ref{fig:BOTuv200L2D2}. The only difference is that at day 10, all fair sample BOT histograms are slightly $\cup$-shaped and biased towards the highest bins, and the same applies for the adjusted sample bot for the largest ensemble size of 64.

  \subsection{Stencils}
  \label{subsB.2}

Again, the matching BOT histograms for the four- and nine-dimensional predictands of 850 hPa temperature on the $2\times 2$ and $3\times 3$ stencils of points, respectively, are rather consistent for all three investigated forecast horizons. Figures \ref{fig:BOTt850dxdyL4D5} and \ref{fig:BOTt850dxdyL4D10} resemble Figure \ref{fig:BOTt850dxdyL4D2}, whereas the fair-, naive-,  and adjusted sample BOT histograms of Figures \ref{fig:BOTt850dxdyL9D5} and \ref{fig:BOTt850dxdyL9D10} are almost identical to the corresponding histograms of Figure \ref{fig:BOTt850dxdyL9D2}.

  \subsection{Vector wind profiles}
\label{subsB.3}

As discussed in Section \ref{subs4.2.3}, the BOT histograms for the 10-dimensional predictand of vector wind on the levels of 200, 300, 500, 700, 850, and 925 hPa behave differently at day 2 (Figure \ref{fig:BOTuv200to925L12D2}) and day 10 (Figure \ref{fig:BOTuv200to925L12D10}). Figure \ref{fig:BOTuv200to925L12D5}, corresponding to day 5, displays some kind of intermediate state; for instance, the transition of the shape of the adjusted BOT histogram for the largest sample size from almost perfectly flat (day 2) to $\cup$-shaped (day 10) seems to be gradual.

\subsection{Verification against analyses}
\label{subsB.4}

Figure \ref{fig:fairBOTallD5}, shows the histograms of the fair sample BOT for all four investigated multivariate predictands for Day 5. In the case of 200 hPa vector wind, the histograms are similar to those at Day~2 (Figure \ref{fig:fairBOTallD2}). The shape of the histogram does not depend on the ensemble size; it is almost flat, with just a tiny overpopulation in the lowest bin. However, for the other three multivariate configurations, the histograms are closer to uniformity than at Day~2 for all ensemble sizes. This increased reliability appears to be mainly due to a lower relative bias of the univariate marginal forecasts at Day~5 where relative bias refers to the mean error of the forecasts normalised by the ensemble standard deviation.

\end{appendix}

\end{document}